\documentclass[conference]{IEEEtran}
\IEEEoverridecommandlockouts
\usepackage{acronym}
\acrodef{pdd}[PDD]{penalty dual decomposition}
\acrodef{admm}[ADMM]{alternating direction method of multipliers}
\acrodef{cadmm}[C-ADMM]{consensus alternating direction method of multipliers}
\acrodef{ps}[PS]{phase shifter}
\acrodef{rfc}[RFC]{radio-frequency chain}
\acrodef{isl}[ISL]{inter-satellite link}
\acrodef{leo}[LEO]{low Earth orbit}
\acrodef{meo}[MEO]{medium Earth orbit}
\acrodef{geo}[GEO]{geostationary Earth orbit}
\acrodef{ut}[UT]{user terminal}
\acrodef{upa}[UPA]{uniform planar array}
\acrodef{csi}[CSI]{channel state information}
\acrodef{ofdm}[OFDM]{orthogonal frequency division multiplexing}
\acrodef{los}[LOS]{line-of-sight}
\acrodef{toa}[TOA]{time-of-arrival}
\acrodef{pace}[PACE]{positioning-aided channel estimation}
\acrodef{mcrb}[MCRB]{misspecified CramÃ©r-Rao bound}
\acrodef{awgn}[AWGN]{additive white Gaussian noise}
\acrodef{crb}[CRB]{CramÃ©r-Rao bound}
\acrodef{cfo}[CFO]{carrier frequency offset}
\acrodef{peb}[PEB]{position error bound}
\acrodef{crb}[CRB]{CramÃ©r-Rao bound}
\acrodef{lb}[LB]{lower bound}
\acrodef{rmse}[RMSE]{root mean squared error}
\acrodef{fim}[FIM]{Fisher information matrix}
\acrodef{tdd}[TDD]{time division duplex} 
\acrodef{wmmse}[WMMSE]{weighted minimal mean squared error}
\acrodef{minlp}[MINLP]{mixed-integer nonlinear programming} 
\acrodef{qcqp}[QCQP]{quadratically constrained quadratic program}
\acrodef{mrt}[MRT]{maximum ratio transmission}
\acrodef{zf}[ZF]{zero-forcing}
\acrodef{sss}[SSS]{single-satellite service}
\acrodef{gnss}[GNSS]{global navigation satellite system}
\acrodef{itu}[ITU]{International Telecommunication Union}
\acrodef{pab}[PAB]{Position-Aided Beamforming}
\acrodef{vdb}[VDB]{Vertically Directed Beamforming}
\acrodef{bs}[BS]{base station}
\acrodef{ntn}[NTN]{non-terrestrial networks}
\acrodef{uav}[UAV]{unmanned aerial vehicle}
\acrodef{tdma}[TDMA]{time-division multiple access}
\acrodef{haps}[HAPS]{high-altitude platform stations}
\acrodef{6g}[6G]{the sixth generation}
\acrodef{bse}[BSE]{beam squint effect}
\acrodef{cp}[CP]{cyclic prefix}
\acrodef{cpu}[CPU]{central processing unit}
\acrodef{dof}[DOF]{degrees-of-freedom}
\acrodef{elaa}[ELAA]{extremely large antenna array}
\acrodef{ff}[FF]{far-field}
\acrodef{las}[L\&S]{localization and sensing}
\acrodef{nf}[NF]{near-field}
\acrodef{ris}[RIS]{reconfigurable intelligent surface}
\acrodef{rtt}[RTT]{round-trip-time}
\acrodef{sinr}[SINR]{signal-to-interference-plus-noise ratio}
\acrodef{sns}[SNS]{spatial non-stationarity}
\acrodef{swm}[SWM]{spherical wave model}

\acrodef{siso}[SISO]{single-input-single-output}
\acrodef{mimo}[MIMO]{multi-input-multi-output}
\acrodef{ue}[UE]{user equipment}
\acrodef{dmimo}[D-MIMO]{distributed MIMO}
\acrodef{sp}[SP]{scatter point}
\acrodef{nlos}[NLOS]{non-line-of-sight}
\acrodef{tdoa}[TDOA]{time-difference-of-arrival}
\acrodef{am}[AM]{artificial multipath}
\acrodef{an}[AN]{artificial noise}
\acrodef{psd}[PSD]{power spectral density}
\acrodef{pdf}[PDF]{probability distribution function}
\acrodef{aoa}[AOA]{angle-of-arrival}
\acrodef{aod}[AOD]{angle-of-departure}
\acrodef{moo}[MOO]{multi-objective optimization}
\acrodef{qos}[QoS]{quality of service}
\acrodef{sdp}[SDP]{semi-definite programming}
\acrodef{lmi}[LMI]{linear matrix inequality}
\acrodef{sdr}[SDR]{semi-definite relaxation}
\acrodef{rcs}[RCS]{radar cross section}
\acrodef{isac}[ISAC]{integrated sensing and communication}
\acrodef{pdd}[PDD]{penalty dual decomposition}
\acrodef{bcd}[BCD]{block coordinate descent}
\acrodef{iui}[IUI]{inter-user interference}
\usepackage{color} 
 
\newcommand{\blue}[1]{{\color{blue}{#1}}}

\usepackage{pgfplots}
\usepackage{tikz}
\usetikzlibrary{calc}
\makeatletter
\newcommand{\gettikzxy}[3]{%
  \tikz@scan@one@point\pgfutil@firstofone#1\relax
  \edef#2{\the\pgf@x}%
  \edef#3{\the\pgf@y}%
}
\usetikzlibrary{spy,backgrounds}
\usepackage{mathrsfs}
\usepackage{booktabs} 
\usepackage{amsmath}
\usepackage{graphicx} 
\usepackage{epstopdf}
\usepackage{amssymb}
\usepackage{amsfonts}
\usepackage{amsthm}
\usepackage{cite}
\usepackage{bm,comment}
\usepackage{algorithm}
\usepackage{algpseudocode}

\usepackage{subeqnarray}
\usepackage{subfigure}
\usepackage{multicol}
\usepackage{multirow}
\usepackage{diagbox}
\usepackage{slashbox}
\usepackage{stfloats}
\usepackage{float}
\usepackage{color} 
\usepackage{cases}
\usepackage{lipsum}

\newtheorem{remark}{\bf{Remark}}
\newtheorem{theorem}{\bf{Theorem}}

\usepackage[bookmarks,colorlinks]{hyperref} 
\hypersetup{colorlinks,citecolor= red,filecolor= blue,linkcolor= blue,urlcolor=blue}
\allowdisplaybreaks
\begin{document}
\setlength{\textfloatsep}{4pt}

\bstctlcite{IEEEexample:BSTcontrol}
\title{Decentralized Cooperative Beamforming for Networked LEO Satellites with Statistical CSI}
\author{
Yuchen Zhang, \emph{Member, IEEE}, Eva Lagunas, \emph{Senior Member, IEEE}, Xue Xian Zheng,\\  Symeon Chatzinotas, \emph{Fellow, IEEE}, and Tareq Y. Al-Naffouri, \emph{Fellow, IEEE}
\thanks{
This publication is based upon work supported by King Abdullah University of Science and Technology (KAUST) under Award No. ORFS-CRG12-2024-6478 and Global Fellowship Program under Award No. RFS-2025-6844.

Yuchen Zhang, Xue Xian Zheng, and Tareq Y. Al-Naffouri are with the Electrical and Computer Engineering Program, Computer, Electrical and Mathematical Sciences and Engineering (CEMSE), King Abdullah University of Science and Technology (KAUST), Thuwal 23955-6900, Kingdom of Saudi Arabia (e-mail: \{yuchen.zhang; 
xuexian.zheng; tareq.alnaffouri\}@kaust.edu.sa).

Eva Lagunas and Symeon Chatzinotas are with the Interdisciplinary Centre for Security, Reliability and Trust (SnT), University of Luxembourg, 1855 Luxembourg City, Luxembourg (e-mail: \{eva.lagunas; Symeon.Chatzinotas\}@uni.lu).
}}
\maketitle

\begin{abstract}
Inter-satellite-link-enabled low-Earth-orbit (LEO) satellite constellations are evolving toward networked architectures that support constellation-level cooperation, enabling multiple satellites to jointly serve user terminals through cooperative beamforming. While such cooperation can substantially enhance link budgets and achievable rates, its practical realization is challenged by the scalability limitations of centralized beamforming designs and the stringent computational and signaling constraints of large LEO constellations.
This paper develops a fully decentralized cooperative beamforming framework for networked LEO satellite downlinks. Using an ergodic-rate-based formulation, we first derive a centralized weighted minimum mean squared error (WMMSE) solution as a performance benchmark. Building on this formulation, we propose a topology-agnostic decentralized beamforming algorithm by \emph{localizing} the benchmark and exchanging a set of globally coupled variables whose dimensions are independent of the antenna number and enforcing consensus over arbitrary connected inter-satellite networks. The resulting algorithm admits fully parallel execution across satellites. To further enhance scalability, we eliminate the consensus-related auxiliary variables in closed form and derive a low-complexity per-satellite update rule that is optimal to local iteration and admits a quasi-closed-form solution via scalar line search.
Simulation results show that the proposed decentralized schemes closely approach centralized performance under practical inter-satellite topologies, while significantly reducing computational complexity and signaling overhead, enabling scalable cooperative beamforming for large LEO constellations.
\end{abstract}


\IEEEpeerreviewmaketitle

\section{Introduction}\label{sec:intro}
\Ac{leo} satellite constellations are rapidly transitioning from isolated access links to \emph{networked} communication infrastructures enabled by \acp{isl}~\cite{halim2021vtm}. This evolution aligns with the 6G vision of ubiquitous connectivity, where \ac{ntn} are expected to complement terrestrial networks and extend coverage to underserved regions~\cite{imt2030vision,ITU2023DATA,6GtakeShape,multiconnect2024cm,eva2025standard,konstantin2025proc}. Compared with \ac{geo}/\ac{meo} systems, \ac{leo} constellations operate at lower altitudes, offering reduced propagation delays and stronger link budgets. Their dense deployments enable multi-satellite coordination, akin to terrestrial standardized multi-connectivity, allowing a \ac{ut} to receive signals over multiple links~\cite{multiconnect2024cm}. In particular, inter-satellite cooperation via \acp{isl} supports coordinated multi-\ac{leo} transmission, alleviating constraints from per-satellite power budgets and finite antenna apertures~\cite{zack2025-DISLAC}.

Motivated by these advantages, networked \ac{leo} cooperative beamforming has attracted growing attention. Early works introduced distributed massive \ac{mimo} concepts over \ac{leo} constellations, demonstrating that satellite cooperation can emulate a virtual large-scale array and yield substantial beamforming gains~\cite{halim2022oj,halim2023oj}. The impact of satellite geometry on throughput has been analyzed in~\cite{Bacci2023taes,luca2025taes}. Other studies exploit \acp{ut}-side spatial processing to facilitate multi-satellite alignment and improve performance~\cite{kexin2024twc}. Position-assisted channel estimation and beamforming leveraging the \ac{los}-dominant nature of \ac{leo} channels have also been explored~\cite{zack2026tcom}, along with joint hybrid beamforming and user scheduling for cooperative satellite networks~\cite{meixia2024twc}. Collectively, these results highlight the potential of networked \ac{leo} cooperation to enhance achievable rates.

Despite this progress, several practical challenges remain. Many cooperative beamforming designs rely on \emph{instantaneous} \ac{csi}~\cite{halim2022oj,halim2021vtm,meixia2024twc,zack2026tcom}, which is difficult to acquire in \ac{leo} systems due to short coherence times, large Doppler shifts, and non-negligible propagation and processing delays~\cite{semiblind2025cl,blockKF2023cl,ming2025twc}. Although recent works mitigate this by exploiting statistical \ac{csi}~\cite{moewin2025jsac,asynLEO2024TWC,yafei2026jsac}, the resulting schemes are typically implemented in an explicitly or implicitly \emph{centralized} manner, aggregating network-wide information at a \ac{cpu} to compute all beamformers jointly, which raises scalability concerns as constellation sizes and user populations grow under stringent on-board constraints.

To reduce complexity, several distributed baselines adopt simple linear beamformers such as \ac{mrt}, \ac{zf}, or their variants~\cite{moewin2023jsac,meixia2024twc,halim2022oj,halim2023oj,Bacci2023taes,luca2025taes}. While computationally efficient, these heuristic methods often incur noticeable performance losses compared with optimization-based designs. More recently, distributed cooperative beamforming approaches based on statistical \ac{csi} have been proposed for \ac{leo} networks~\cite{zack2025twc}. However, these methods are typically restricted to specific \ac{isl} topologies and struggle to accommodate the diverse variation of \ac{isl} connectivity in practice~\cite{halim2021vtm}. Moreover, scalability remains an issue, as existing solutions often rely on sequential execution or centralized information fusion.

\begin{figure}[t]
		\centering
		\includegraphics[width=1.0 \linewidth]{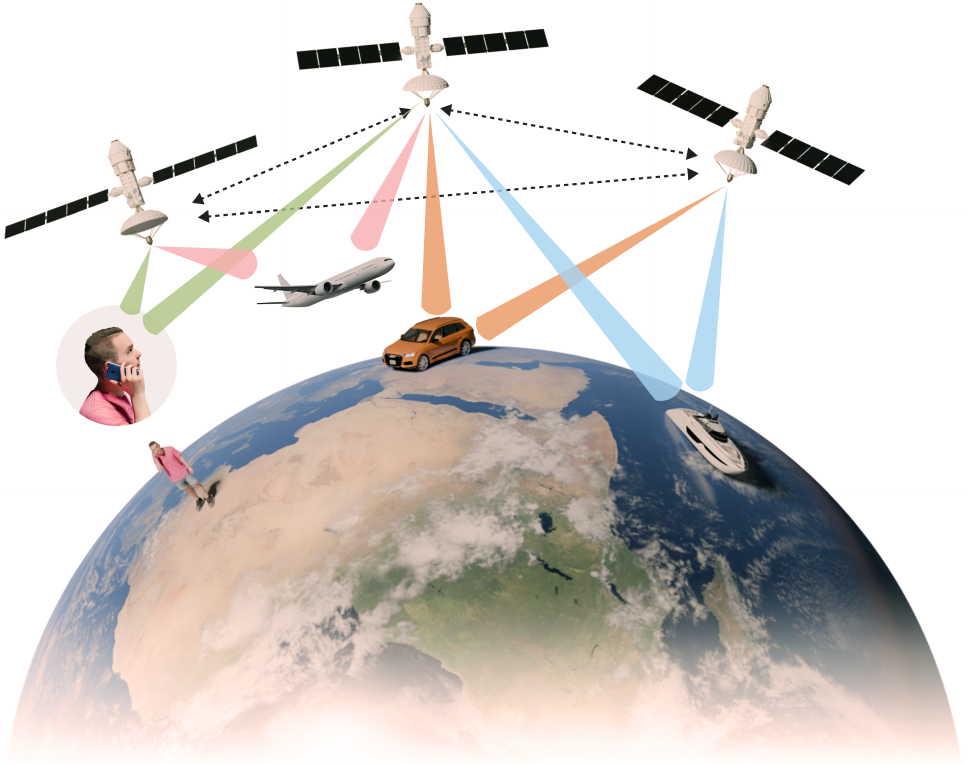}
		\caption{An illustration of networked-LEO satellite system, where multiple LEO satellites collaboratively serve UTs through cooperative beamforming.}
		\label{sys_mod}
\end{figure}

This paper aims to bridge the above gaps by developing a \emph{decentralized} cooperative beamforming framework for networked \ac{leo} satellites that (i) avoids reliance on instantaneous \ac{csi}, (ii) is agnostic to the underlying \ac{isl} topology, and (iii) scales to large constellations through \emph{fully parallel} per-satellite execution with manageable local computational complexity and network-wide signaling overhead. The main contributions of this paper are summarized as follows:
\begin{itemize}
    \item \textbf{Statistical-\ac{csi}-based cooperative beamforming formulation:}
    We consider a networked \ac{leo} downlink in which multiple satellites jointly serve multiple \acp{ut} via cooperative beamforming over \acp{isl}. Motivated by the short coherence time and long propagation delays of \ac{leo} links, which make instantaneous \ac{csi} acquisition unreliable across cooperating satellites, we adopt a statistical-\ac{csi}-based metric and formulate a per-satellite power-constrained sum-rate maximization problem using a hardening-bound-based ergodic rate lower bound.

    \item \textbf{Topology-agnostic and fully parallel decentralized design:}
    Building on a centralized \ac{wmmse}-based formulation, we develop a topology-agnostic decentralized cooperative beamforming framework by localizing a carefully selected set of globally coupled variables whose dimensions are \emph{independent} of the antenna number, a property essential for alleviating \ac{isl} overhead. Network-wide consistency is enforced via \ac{cadmm} over an arbitrary connected \ac{isl} graph. Importantly, the resulting algorithm admits \emph{fully parallel execution} across all satellites, making it applicable to representative Ring, Star, and Mesh topologies as well as \emph{general connected \ac{leo} networks}, with guaranteed convergence. Topology-agnosticism is essential to adapt to dynamic \ac{leo} \ac{isl} re-formation, requiring no re-derivation when the cooperating graph changes.

    \item \textbf{Low-complexity per-satellite solution:}
    To fit the tight on-board compute budgets of \ac{leo} satellites, we derive closed-form expressions for the consensus auxiliaries and eliminate them from the local optimization, yielding an equivalent per-satellite problem involving solely local beamformers. Through strong duality and the problem's eigen-structure, we develop a \emph{quasi-closed-form} solution based on scalar line search, avoiding generic solvers and enabling efficient, scalable decentralized implementation.

    \item \textbf{Comprehensive evaluation and practical insights:}
    Extensive simulations demonstrate that the proposed decentralized schemes achieve sum-rate performance close to that of the centralized benchmark under practical \ac{isl} topologies, while significantly reducing signaling overhead and achieving orders-of-magnitude lower runtime. The results further reveal that dense \ac{isl} is not required to reap most of the cooperative gains, validating the scalability of the proposed framework. 
\end{itemize}

The remainder of this paper is organized as follows.
Section~\ref{sec_system} introduces the system and channel models and formulates the statistical-\ac{csi}-based cooperative beamforming problem.
Section~\ref{sec_central} presents the centralized \ac{wmmse}-based benchmark solution.
Section~\ref{sec_decentral} develops the decentralized cooperative beamforming framework over arbitrary connected \ac{isl} topologies, enabled by a strategical combination of \ac{wmmse} and \ac{cadmm}.
Section~\ref{sec_lc_scheme} proposes the low-complexity optimal local solver.
Finally, Section~\ref{sec_numer} provides numerical results, and Section~\ref{sec_con} concludes the paper.

\emph{Notations:}
Lowercase letters denote scalars, whereas bold lowercase and bold uppercase letters represent vectors and matrices, respectively. The Euclidean norm of a vector $\boldsymbol{a}$ is denoted by $\|\boldsymbol{a}\|$, and the Frobenius norm of a matrix $\boldsymbol{A}$ is denoted by $\|\boldsymbol{A}\|_{\mathrm{F}}$. The operators $(\cdot)^*$, $(\cdot)^{\mathsf{T}}$, and $(\cdot)^{\mathsf{H}}$ correspond to complex conjugation, transpose, and Hermitian transpose, respectively. The real part of a complex scalar $a$ is written as $\Re\{a\}$. The symbols $\mathbb{E}[\cdot]$, $\mathbb{V}[\cdot]$, and $\mathrm{diag}[\cdot]$ denote the expectation, variance, and diagonalization operators, respectively. A circularly symmetric complex Gaussian random vector with mean $\boldsymbol{\mu}$ and covariance matrix $\boldsymbol{C}$ is denoted by $\mathcal{CN}(\boldsymbol{\mu}, \boldsymbol{C})$. Finally, $\boldsymbol{0}_N$ and $\boldsymbol{0}_{N \times N}$ represent the all-zero vector of length $N$ and the $N \times N$ zero matrix, respectively, while $\boldsymbol{I}_N$ denotes the $N$-dimensional identity matrix.

\section{System Model}\label{sec_system}
As illustrated in Fig.~\ref{sys_mod}, we consider a networked-\ac{leo} satellite system where $S$ \ac{leo} satellites jointly serve $U$ \acp{ut} in the downlink. Each \ac{ut} is equipped with a single omnidirectional antenna, while each satellite employs a \ac{upa} with $N = N_{\text{h}} N_{\text{v}}$ half-wavelength-spaced elements, where $N_{\text{h}}$ and $N_{\text{v}}$ denote the numbers of antennas along the horizontal and vertical dimensions, respectively. For simplicity, all satellites are assumed to share the same array configuration, though extending the model to heterogeneous arrays is straightforward. Leveraging \acp{isl}, regenerative satellites can exchange \ac{ut} information, enabling cooperative downlink transmission in which each \ac{ut} may be served by multiple satellites.

\subsection{Channel Model}
Consider the downlink transmission from the $s$-th satellite to the $u$-th \ac{ut}. Let $f$ and $t$ denote the signal frequency and time instant, respectively. The channel is expressed as
\begin{align}\label{chan_mod} 
\boldsymbol{h}_{s,u}\left(t,f \right) = &\sum_{m=0}^{M_{s,u}}\alpha_{s,u,m} G\left(\theta_{s,u,m}^{\mathrm{el}}\right)e^{\jmath 2 \pi\left(t \upsilon_{s,u,m} - f \tau_{s,u,m} \right) } \notag \\
&\times \boldsymbol{a}\left(\boldsymbol{\theta}_{s,u,m}\right),    
\end{align}  
where $M_{s,u}$ is the number of propagation paths and $\alpha_{s,u,m}$ denotes the complex gain of the $m$-th path. The path with index $m=0$ corresponds to the \ac{los} component, while the remaining ones are \ac{nlos}. The parameters $\tau_{s,u,m}$ and $\upsilon_{s,u,m}$ represent the propagation delay and Doppler shift, respectively. 

The satellite array response vector is denoted by $\boldsymbol{a}(\boldsymbol{\theta}_{s,u,m}) \in \mathbb{C}^{N}$, where $\boldsymbol{\theta}_{s,u,m} = [\theta_{s,u,m}^{\mathrm{az}},\,\theta_{s,u,m}^{\mathrm{el}}]^{\mathsf{T}}$ collects the \ac{aod} (azimuth and elevation). The antenna radiation pattern $G(\theta_{s,u,m}^{\mathrm{el}})$ depends only on the elevation angle and is boresight-symmetric~\cite{balanis2005antenna,zack2026tcom,zack2025twc}. Without loss of generality, the \ac{upa} at each satellite lies on the local XY-plane of a right-handed coordinate system. Define $\boldsymbol{n}(N) = [0,\ldots,N-1]^{\mathsf{T}}$. The steering vector is expressed as
\begin{equation}  
\boldsymbol{a}\left(\boldsymbol{\theta}_{s,u,m}\right) = e^{-\jmath 2 \pi \phi_{s,u,m}^{\text{h}} \boldsymbol{n}\left(N_{\text{h}}\right)} \otimes e^{-\jmath 2 \pi \phi_{s,u,m}^{\text{v}} \boldsymbol{n}\left(N_{\text{v}}\right)},  
\end{equation}  
where $\phi_{s,u,m}^{\text{h}} = d \cos \theta_{s,u,m}^{\mathrm{az}} \cos \theta_{s,u,m}^{\mathrm{el}}/\lambda$ and $\phi_{s,u,m}^{\text{v}} = d \sin \theta_{s,u,m}^{\mathrm{az}} \cos \theta_{s,u,m}^{\mathrm{el}}/\lambda$. Here, $d$ denotes the antenna spacing and $\lambda$ is the wavelength corresponding to the carrier frequency.

In \ac{leo} satellite systems, the satellite altitude is much larger than the typical scatterer distribution radius near the \ac{ut}. Therefore, the \acp{aod} in \eqref{chan_mod} can be approximated as identical for all paths, i.e., $\boldsymbol{\theta}_{s,u,m} \approx \boldsymbol{\theta}_{s,u}, \forall m$. Similarly, the Doppler shift of each path can be decomposed as $\upsilon_{s,u,m} = \upsilon_{s,u,m}^{\text{Sat}} + \upsilon_{s,u,m}^{\text{UT}}$, where $\upsilon_{s,u,m}^{\text{Sat}}$ and $\upsilon_{s,u,m}^{\text{UT}}$ are induced by the satellite and the \ac{ut}, respectively. Since the satellite velocity dominates and is nearly identical for all rays, we can approximate $\upsilon_{s,u,m}^{\text{Sat}} \approx \upsilon_{s,u}^{\text{Sat}}, \forall m$.

Let $\tau_{s,u} = \tau_{s,u,0}$ denote the minimum (LOS) delay, and define the differential delay $\tau_{s,u,m}^{\text{Diff}} = \tau_{s,u,m} - \tau_{s,u}$. Substituting these relations into \eqref{chan_mod}, the equivalent LOS channel can be represented as
\begin{equation}\label{chan_mod_approx} 
\boldsymbol{h}_{s,u}\left(t,f \right) = \alpha_{s,u} e^{\jmath 2 \pi\left(t \upsilon_{s,u}^{\text{Sat}} - f \tau_{s,u} \right) }\boldsymbol{b}\left(\boldsymbol{\theta}_{s,u}\right),
\end{equation}
where $\boldsymbol{b}(\boldsymbol{\theta}_{s,u}) = G(\theta_{s,u}^{\mathrm{el}})\boldsymbol{a}(\boldsymbol{\theta}_{s,u})$ and 
\begin{equation*}
\alpha_{s,u} = \sum_{m=0}^{M_{s,u}}\alpha_{s,u,m}e^{\jmath 2 \pi\left(t\upsilon_{s,u,m}^{\text{UT}} - f \tau_{s,u,m}^{\text{Diff}} \right) }
\end{equation*}
denotes the composite channel gain. The random variable $\alpha_{s,u}$ follows a Rician distribution with factor $\kappa_{s,u}$ and mean power $\mathbb{E}[|\alpha_{s,u}|^2] = \gamma_{s,u}$~\cite{poor2024tsp,you2020jsac}.

Physically, $\alpha_{s,u}$ captures the residual frequency and time variation due to user mobility and multipath. Under the slow-\ac{ut}-mobility and narrowband assumptions, $\alpha_{s,u}$ evolves slowly in time and remains nearly flat in frequency\cite{zack2025twc}, so its explicit $(t,f)$ dependence is omitted. Accordingly, $\alpha_{s,u}\sim\mathcal{CN}(\bar{\alpha}_{s,u},\beta_{s,u})$ with \ac{los} mean $\bar{\alpha}_{s,u} = \sqrt{\kappa_{s,u}\gamma_{s,u}/(1+\kappa_{s,u})}$ and scatter variance $\beta_{s,u} = \gamma_{s,u}/(1+\kappa_{s,u})$. We treat $\bar{\alpha}_{s,u}$ as real and non-negative because the deterministic \ac{los} phase $-2\pi f \tau_{s,u}$ is removed by the position-driven precompensation introduced below, an operating regime adopted in statistical-\ac{csi} cooperative \ac{leo} precoding\cite{you2020jsac,moewin2025jsac,yafei2026jsac,zack2025twc} and made feasible in practice given dedicated phase-synchronization techniques studied in\cite{yafei2026sync,liz2022access}.

Finally, since the satellite's position and velocity are accurately predictable, both Doppler and delay effects can be jointly compensated as shown in~\cite{you2020jsac,kexin2024twc,poor2024tsp,moewin2025jsac}. Assuming perfect compensation, the channel in~\eqref{chan_mod_approx} simplifies to
\begin{equation}
\boldsymbol{h}_{s,u} = \alpha_{s,u}\boldsymbol{b}\left(\boldsymbol{\theta}_{s,u}\right).
\end{equation}
Imperfect Doppler/delay compensation introduces a residual random phase that erodes coherent gain, explicitly characterized in dedicated robustness analyses\cite{asynLEO2024TWC,shiyu2025tvt}. Our framework still applies in such regimes by replacing $\bar{\alpha}_{s,u}$ with its residual-attenuated counterpart, with no algorithmic change.

\subsection{Signal Model}
Let $\boldsymbol{s}[\ell]= [s_1[\ell],\ldots,s_U[\ell]]^{\mathsf{T}}\sim \mathcal{CN}(\boldsymbol{0}_U,\boldsymbol{I}_U)$ denote the collection of data streams for $U$ \acp{ut} during the $\ell$-th symbol. The corresponding transmit signal at the $s$-th \ac{leo} satellite is expressed as
\begin{equation}
\boldsymbol{x}_s\left[\ell\right] = \boldsymbol{W}_s\mathrm{diag}\left(\boldsymbol{\delta}_s\right)\boldsymbol{s}\left[\ell\right],
\end{equation}
where $\boldsymbol{W}_s = [\boldsymbol{w}_{s,1},\ldots,\boldsymbol{w}_{s,U}] \in \mathbb{C}^{N \times U}$ represents the beamformer, and $\boldsymbol{\delta}_s = [\delta_{s,1},\ldots,\delta_{s,U}]^{\mathsf{T}}$ denotes the scheduler, with each element being either $0$ or $1$. Specifically, the $u$-th \ac{ut} is served via the $s$-th satellite if $\delta_{s,u} = 1$, and is not if $\delta_{s,u} = 0$.

The signal received at the $u$-th \ac{ut} during the $\ell$-th symbol is given by  
\begin{align}\label{receive_sig}
y_u\left[\ell\right] &= \sum_{s=1}^{S} \boldsymbol{h}_{s,u}^{\mathsf{T}}   \boldsymbol{W}_s\mathrm{diag}\left(\boldsymbol{\delta}_s\right)\boldsymbol{s}\left[\ell\right] + n_u\left[\ell\right]  \\
&= \sum_{s=1}^{S} \boldsymbol{h}^{\mathsf{T}}_{s,u} \delta_{s,u}\boldsymbol{w}_{s,u} s_u\left[\ell\right]   + \underbrace{\sum_{l\neq u}^{U} \sum_{s=1}^{S} \boldsymbol{h}^{\mathsf{T}}_{s,u}\delta_{s,l} \boldsymbol{w}_{s,l} s_l\left[\ell\right]}_{\text{IUI}}\notag \\
&\quad + n_u\left[\ell\right],\notag
\end{align}  
where $n_u[\ell]\sim \mathcal{CN}(0,\sigma^2)$ denotes the \ac{awgn} with noise variance given by $\sigma^2 = N_0 B$. Here, $N_0$ is the single-sided \ac{psd} and $B$ denotes the signal bandwidth.

\subsection{Problem Formulation}
Since obtaining accurate instantaneous \ac{csi} in \ac{leo} satellite systems is hindered by short coherence time and long propagation delay relative to terrestrial links \cite{kexin2023twc,chang2023iotj,semiblind2025cl,blockKF2023cl,ming2025twc}, we focus on a statistical performance metric, namely the ergodic rate, instead of its instantaneous counterpart. To ensure analytical tractability, we approximate the ergodic rate using its lower bound, known as the hardening bound \cite{Marzetta2016mMIMO,caire2018twc}.

Let $\boldsymbol{g}_{u,l} = [g_{1,u,l},\ldots,g_{S,u,l}]^{\mathsf{T}}$ with $g_{s,u,l} = \boldsymbol{b}^{\mathsf{T}}(\boldsymbol{\theta}_{s,u}) \delta_{s,l}\boldsymbol{w}_{s,l}$ and $\boldsymbol{\alpha}_u = [\alpha_{1,u},\ldots,\alpha_{S,u}]^{\mathsf{T}}$. Based on \eqref{receive_sig}, the ergodic rate lower bound of the $u$-th \ac{ut} can be expressed as 
\begin{equation}\label{rate_lb}
R_u^{\text{LB}} = \log_2 \left(1 + \frac{\left|\mathbb{E}\left[\Gamma_{u,u}\right]\right|^2}{\mathbb{V}\left[\Gamma_{u,u}\right] + \sum_{l\ne u}^{U}\mathbb{E}\left[ \left|\Gamma_{u,l}\right|^2 \right] + \sigma^2}\right),   
\end{equation}
where $\Gamma_{u,l}=\sum_{s=1}^{S} \alpha_{s,u} g_{s,u,l}$. It can be derived that 
\begin{subequations}\label{rate_component}
\begin{align}
\mathbb{E}\left[\Gamma_{u,u}\right] & = \sum_{s=1}^{S}\bar{\alpha}_{s,u} g_{s,u,u},\\
\mathbb{V}\left[\Gamma_{u,u}\right] & = \sum_{s=1}^{S}\beta_{s,u} \left|g_{s,u,u}\right|^2, \\
\mathbb{E}\left[ \left|\Gamma_{u,l}\right|^2 \right] & = \boldsymbol{g}_{u,l}^{\mathsf{H}} \boldsymbol{T}_u \boldsymbol{g}_{u,l},
\end{align}
\end{subequations}
where $\boldsymbol{T}_u = \mathbb{E}\left[\boldsymbol{\alpha}_u \boldsymbol{\alpha}_u^{\blue{\mathsf{H}}}\right] \in \mathbb{C}^{S \times S}$ denotes the inter-satellite channel gain correlation for user $u$. Owing to the large inter-satellite spacing, $\alpha_{i,u}$ and $\alpha_{j,u}$ are assumed independent for $i\neq j$, leading to
\begin{equation}\label{T_parameter}
\boldsymbol{T}_u= \bar{\boldsymbol{\alpha}}_u \bar{\boldsymbol{\alpha}}_u^{\mathsf{T}} + \mathrm{diag}\left(\boldsymbol{\beta}_u \right),
\end{equation}
where $\bar{\boldsymbol{\alpha}}_u  = [\bar{\alpha}_{1,u},\ldots,\bar{\alpha}_{S,u}]^{\mathsf{T}}$ and $\boldsymbol{\beta}_u  = [\beta_{1,u},\ldots,\beta_{S,u}]^{\mathsf{T}}$.
As shown in \eqref{rate_component}, the resulting communication rate thus depends solely on the \emph{statistical channel parameters} rather than the instantaneous \ac{csi}.

\begin{remark}\label{sched_decouple_remark}
Jointly optimizing $\boldsymbol{\delta}_s$ and $\boldsymbol{W}_s$ would yield a \ac{minlp}, which is ill-suited to the on-board computational and power budgets of \ac{leo} satellites. We therefore decouple the two procedures, in line with prior \ac{leo} designs that take the scheduled user set as given\cite{you2020jsac,zack2025twc}: $\boldsymbol{\delta}_s$ is fixed first by a geometry-driven heuristic, after which $\boldsymbol{W}_s$ is optimized. Several scheduling rules are evaluated in Section~\ref{sec_numer} to characterize how the choice of scheduler shapes the achievable cooperative gain. A joint design is left to future work.
\end{remark}

In what follows, we jointly optimize the beamformers across all \ac{leo} satellites, i.e., $\boldsymbol{W}_s,\forall s$, to maximize the network sum rate, while assuming that the binary scheduling variables $\boldsymbol{\delta}_s,\forall s$ are pre-determined by a geometry-driven heuristic, as discussed in Remark~\ref{sched_decouple_remark}.
Under per-satellite power budgets, the sum-rate maximization problem is formulated as
\begin{subequations}\label{prob_ori}
\begin{align}
\mathop {\max }\limits_{ \{\boldsymbol{w}_{s,u}\}_{\forall s, u}} \;\;\; &\sum_{u=1}^{U} R_u^{\text{LB}}  \\
{\rm{s.t.}}\;\;\;
&\sum_{u = 1}^{U} \delta_{s,u}
\left\| \boldsymbol{w}_{s,u} \right\|^2  \le P_s, \;\forall s,\label{prob_ori_pow_bud}
\end{align}  
\end{subequations}
where $P_s$ is the power budget of the $s$-th \ac{leo}.

Due to the coupling among the beamformers of different satellites in the objective function, problem~\eqref{prob_ori} is non-convex and cannot be solved directly. To address this challenge, we first develop a centralized cooperative beamforming scheme to tackle~\eqref{prob_ori}. This centralized formulation not only provides a performance upper bound but also establishes the foundation for the subsequent decentralized designs.

\begin{remark}
Due to the presence of binary scheduling variables, the beamformer from each satellite to a given \ac{ut} becomes effective only under non-zero scheduler. Let $\mathcal{U}_s$ denote the set of \acp{ut} scheduled by the $s$-th satellite, i.e., $\delta_{s,u}=1$ for all $u \in \mathcal{U}_s$ and $\delta_{s,u}=0$ for all $u \notin \mathcal{U}_s$. Consequently, for the $s$-th \ac{leo} satellite, it suffices to design the beamformers $\boldsymbol{w}_{s,u}$ only for $u \in \mathcal{U}_s$, since the remaining beamformers are inactive and do not contribute to the transmission.
\end{remark}

\section{Centralized Cooperative Beamforming Design}\label{sec_central}
In this section, we develop a centralized cooperative beamforming optimization framework to solve \eqref{prob_ori}. In this scheme, the optimization is performed at a \ac{cpu}, for instance, located at a master or central satellite, which collectively determines the beamformers before disseminating the results to all participating \ac{leo} satellites. To facilitate the solution, we employ the \ac{wmmse} framework to handle the fractional \ac{sinr} expression in $R_u^{\text{LB}}$. 
Based on the transformations, we propose an iterative optimization procedure to efficiently solve \eqref{prob_ori}.

\subsection{WMMSE-Based Framework}

To eliminate the fractional structure of the \ac{sinr} term in \eqref{rate_lb} and enable tractable optimization, we adopt the \ac{wmmse} framework~\cite{shi2011wmmse,zack2026tcom}. This approach introduces auxiliary variables $\mu_{u}$ and $\nu_{u}$, transforming the original problem \eqref{prob_ori} into an equivalent fractional-free formulation expressed as
\begin{subequations}\label{prob_refor}
\begin{align}
\mathop {\min }\limits_{\{\mu_{u},\nu_{u}\}_{\forall u},\{\boldsymbol{w}_{s,u}\}_{\forall s, u\in \mathcal{U}_s}}
&\sum_{u=1}^{U} \left(\nu_{u} \Upsilon_{u} -  \ln \nu_{u}\right) \\
\text{{\rm s.t.}} \;\;\;
&\eqref{prob_ori_pow_bud},
\end{align}
\end{subequations}
where $\Upsilon_{u} = 
|1 - \mu_{u}\sum_{s=1}^{S}\bar{\alpha}_{s,u} g_{s,u,u}|^2 + |\mu_{u}|^2\Psi_u$. 
Here, we have $\Psi_u =  \sum_{s=1}^{S}\beta_{s,u}|g_{s,u,u}|^2
+ \sum_{l\ne u}^{U}\boldsymbol{g}_{u,l}^{\mathsf{H}}\boldsymbol{T}_u\boldsymbol{g}_{u,l} + \sigma^2$.

With the above reformulation, the optimization can be carried out iteratively by updating the involved variables through a sequence of tractable subproblems.

\emph{1) Update of $\mu_{u}$:}
For fixed $\nu_{u}$ and $\boldsymbol{w}_{s,u}$, the optimal $\mu_{u}$ is obtained by minimizing $\Upsilon_{u}$ with respect to $\mu_{u}$, i.e., by setting $\partial \Upsilon_{u} / \partial \mu_{u} = 0$. The resulting closed-form solution is
\begin{equation}\label{update_mu}
\mu_{u} = 
\frac{\left(\sum_{s=1}^{S}\bar{\alpha}_{s,u} g_{s,u,u}\right)^{\mathsf{*}}}
{\left|\sum_{s=1}^{S}\bar{\alpha}_{s,u} g_{s,u,u}\right|^2 + \Psi_u},\; \forall u.
\end{equation}

\emph{2) Update of $\nu_{u}$:}
Given $\mu_{u}$ and $\boldsymbol{w}_{s,u}$, the optimal $\nu_{u}$ that maximizes the objective in \eqref{prob_refor} is expressed as
\begin{equation}\label{update_nu}
\nu_{u} = \frac{1}{\Upsilon_{u}},\; \forall u.
\end{equation}

\emph{3) Update of beamformer $\boldsymbol{w}_{s,u}$:}
With $\mu_{u}$ and $\nu_{u}$ fixed, the beamformer $\boldsymbol{w}_{s,u}$ can be optimized by solving
\begin{subequations}\label{cent_bf_opt}
\begin{align}
\mathop {\min }\limits_{\{\boldsymbol{w}_{s,u}\}_{\forall s, u\in \mathcal{U}_s }} \;\;\;
&\sum_{u=1}^{U} \nu_{u}\Upsilon_{u}  \\
\text{{\rm s.t.}} \;\;\;
&\eqref{prob_ori_pow_bud}.
\end{align}
\end{subequations}
Since $\Upsilon_{u}$ is a convex quadratic function with respect to $\boldsymbol{w}_{s,u}$, the optimization problem in \eqref{cent_bf_opt} constitutes a convex \ac{qcqp}, which can be efficiently solved using standard convex optimization toolboxes, such as CVX.

\renewcommand{\algorithmicrequire}{\textbf{Input:}}
\renewcommand{\algorithmicensure}{\textbf{Output:}}
\begin{algorithm}[t]
\caption{Centralized Cooperative Beamforming Design for Networked \ac{leo} Satellites}
\label{wmmse_cent}
\begin{algorithmic}[1]
\State \textbf{Initialize}: {$\boldsymbol{W}_s, \forall s$, 
$\boldsymbol{\delta}_s,\forall s$;}
\Repeat
\State {Update $\mu_{u}, \forall u$ using \eqref{update_mu};}
\State {Update $\nu_{u}, \forall u$ using \eqref{update_nu};}
\State {Update $\boldsymbol{w}_{s,u}, \forall s, u\in \mathcal{U}_s$ by solving \eqref{cent_bf_opt} via CVX;}
\Until {the relative reduction in the objective value falls below a predefined threshold or a maximum number of iterations is reached;}
\State \textbf{Output}: {$\boldsymbol{w}_{s,u}, \forall s, u\in \mathcal{U}_s$.}
\end{algorithmic}
\end{algorithm}

\subsection{Convergence and Complexity}
The overall procedure for solving \eqref{prob_ori} is summarized in Algorithm~\ref{wmmse_cent}. The convergence of Algorithm~\ref{wmmse_cent} can be readily established since each iteration monotonically increases the objective value of \eqref{prob_refor}, which is upper-bounded due to the finite power budget. The computational complexity per iteration is dominated by solving \eqref{cent_bf_opt}, which, as a \ac{qcqp} problem, entails computational cost on the order of $\mathcal{O}((N\sum_{s=1}^{S}|\mathcal{U}_s|)^3)$. Here, $\left|\mathcal{U}_s\right|$ denotes the cardinality of $\mathcal{U}_s$.

\begin{remark}\label{cent_limit}
Future \ac{leo} satellite systems are anticipated to employ massive antenna arrays with a large number of elements~\cite{you2020jsac,you2024twc,kexin2023twc,kexin2024twc}. In such scenarios, the centralized cooperative beamforming design imposes a substantial computational burden on the \ac{cpu}, particularly when a large number of satellites participate in cooperative transmission and/or many \acp{ut} are scheduled for service.
Unlike terrestrial \acp{bs}, which typically operate with relatively abundant processing power and energy budgets, \ac{leo} satellites face stringent constraints on both computational capability and on-board resources. These limitations make large-scale centralized processing impractical. To accommodate such constraints and enable scalable networked \ac{leo} cooperative beamforming, it is therefore desirable to develop decentralized optimization schemes that distribute the computational workload across satellites, thereby improving robustness, efficiency, and scalability of the overall system.
\end{remark}

\subsection{LEO Satellite Network Topologies}

\begin{figure}[t]
\centering
\includegraphics[width=1\linewidth]{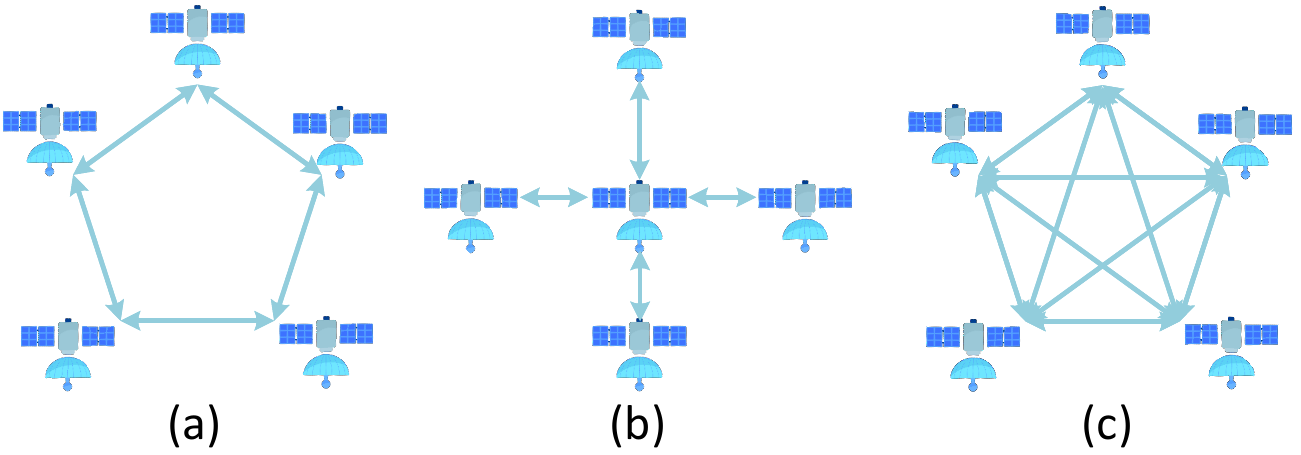}
\caption{Representative \ac{isl} topologies: (a) Ring; (b) Star; and (c) Mesh.}\label{leo_topo}
\end{figure}

Unlike terrestrial \ac{bs} networks, which are typically interconnected through fixed and largely static backhaul infrastructures, \ac{leo} satellite networks rely on dynamic \acp{isl} whose connectivity evolves over time due to satellite mobility and the varying availability of neighboring nodes. As illustrated in Fig. \ref{leo_topo}, several representative \ac{isl} topologies can be considered, including the \emph{Ring}, \emph{Star}, and \emph{Mesh} configurations. In the Ring topology, each satellite maintains links with exactly two neighboring satellites. In the Star topology, a single central satellite connects to all peripheral satellites, while no direct links exist among the peripheral nodes themselves. In the Mesh topology, every satellite is directly connected to all others through dedicated \acp{isl}. 

In practice, the actual \ac{isl} connectivity often appears as a hybrid of these canonical structures and can be modeled using a graph $\mathcal{G}=(\mathcal{V},\mathcal{E})$, where $\mathcal{V}$ and $\mathcal{E}$ represent the sets of satellites and inter-satellite links, respectively. As long as $\mathcal{G}$ is \emph{connected}, any pair of satellites can exchange information, either directly or indirectly, thereby ensuring that local updates can propagate throughout the network and influence the global optimization process. In this paper, we develop a decentralized cooperative beamforming optimization framework that accommodates \emph{arbitrary connected \ac{leo} network topologies} over $\mathcal{G}$, thus providing both scalability and broad applicability.

The topology-agnostic structure aligns naturally with the dynamic nature of \ac{leo} \ac{isl} connectivity, which evolves over time due to orbital motion, handovers, and link blockages\cite{halim2021vtm,liz2022access}, in contrast to the largely static backhaul of terrestrial cooperative networks. In practice, when the underlying graph $\mathcal{G}$ evolves and varies over time, only each satellite's local neighbor set $\mathcal{G}_s$ and the associated consensus exchange need to be refreshed, without altering algorithmic applicability.

\section{Decentralized Cooperative Beamforming Design}\label{sec_decentral}

In this section, we tailor the \ac{wmmse} framework to make it compatible with the \ac{cadmm}\cite{gonzalo2010tsp,chang2015tsp}, thereby enabling the development of a decentralized cooperative beamforming scheme applicable to any connected \ac{leo} satellite network topology. In the proposed scheme, all participating \ac{leo} satellites perform local signal processing and optimization in parallel. After each local update, intermediate parameters are exchanged bidirectionally among neighboring satellites according to the underlying \ac{isl} topology. This iterative exchange continues until network-wide convergence is achieved.

\subsection{Decentralization via C-ADMM}
Before proceeding to the decentralization process, we note that the statistical channel parameters, i.e., $\bar{\alpha}_{s,u}, \forall s,u$ and $\beta_{s,u}, \forall s,u$ (with $\boldsymbol{T}_u, \forall u$ inferred directly from \eqref{T_parameter}), must be globally available across the network, as they are required for local processing at each satellite. This requirement introduces a certain amount of communication overhead during initialization. However, since the proposed optimization framework depends only on statistical channel parameters, which evolve much more slowly than their instantaneous counterparts, the update frequency of these parameters is relatively low. Therefore, the signaling overhead associated with collecting and distributing these parameters across the network is negligible and thus omitted for simplicity.

Solving \eqref{prob_ori} using Algorithm~\ref{wmmse_cent} requires a centralized \ac{cpu}, primarily because the beamformers across different satellites, embedded in the terms $g_{s,u,l}, \forall s,u,l$, are mutually coupled through $\Gamma_u, \forall u$. As a result, each satellite \emph{cannot independently update} its local beamformers without accessing the quantities $g_{i,u,l}, \forall i \neq s, u, l$. To overcome this limitation, we \emph{localize} the optimization at each satellite by introducing local copies of the global variables $\boldsymbol{g}_{u,l}$, denoted by $\boldsymbol{g}_{u,l}^{(s)}$. For the $s$-th satellite, the local copy corresponding to its own beamformer, i.e., the $s$-th entry of $\boldsymbol{g}_{u,l}^{(s)}$, naturally satisfies
\begin{equation}\label{local_bf_constraint}
g_{s,u,l}^{(s)} = \boldsymbol{b}^{\mathsf{T}}\left(\boldsymbol{\theta}_{s,u}\right)\, \delta_{s,l}\, \boldsymbol{w}_{s,l},\;\forall u,l.    
\end{equation}

Then, each satellite steers the network toward agreement among all local copies of $g_{i,u,l}$ by enforcing consensus between its locally maintained variables $g_{i,u,l}^{(s)}, \forall i \neq s$, and the corresponding copies received from its neighbors (including its own self-reference), denoted by $\tilde{g}_{i,u,l}^{(j)}, \forall i \neq s,\, j \in \mathcal{G}_s \cup \{s\}$. Here, $\mathcal{G}_s$ represents the set of satellites directly connected to the $s$-th satellite via \acp{isl}. The tilde notation is introduced to emphasize that these quantities are treated as fixed copies rather than optimization variables.
 
Note that, due to the presence of the binary scheduler, which appears inherently as a multiplicative factor, we have $g_{i,u,l}^{(s)} = 0, \forall i \ne s, u,l \notin \mathcal{G}_i$. Therefore, it should be emphasized that only the variables $g_{i,u,l}^{(s)}, \forall i \ne s, u, l \in \mathcal{G}_i$, need to be included as optimization variables.
For notational compactness, we define  
$\boldsymbol{g}_{-s,u,l}^{(s)} = [g_{1,u,l}^{(s)},\ldots,g_{s-1,u,l}^{(s)},g_{s+1,u,l}^{(s)},\ldots,g_{S,u,l}^{(s)}]^{\mathsf{T}}$ and $\tilde{\boldsymbol{g}}_{-s,u,l}^{(j)} = [\tilde{g}_{1,u,l}^{(j)},\ldots,\tilde{g}_{s-1,u,l}^{(j)},\tilde{g}_{s+1,u,l}^{(j)},\ldots,\tilde{g}_{S,u,l}^{(j)}]^{\mathsf{T}}$, which collect $g_{i,u,l}^{(s)}, \forall i \ne s,$ and $\tilde{g}_{i,u,l}^{(j)}, \forall i \ne s$.   
The localized version of \eqref{prob_refor} at the $s$-th satellite can be formulated as

\begin{subequations}\label{prob_refor_consensus}
\begin{align}
&\mathop {\min }\limits_{\substack{\{\mu_{u}^{(s)},\nu_{u}^{(s)}\}_{\forall u},\{\boldsymbol{w}_{s,u}\}_{\forall u\in \mathcal{U}_s},\\\{g_{i,u,l}^{(s)}\}_{\forall i \ne s,u,l \in \mathcal{U}_i}}}  
\sum_{u=1}^{U} \left(\nu_{u}^{(s)} \Upsilon_{u}^{(s)} - \ln \nu_{u}^{(s)}\right) \\
&\;\;\;\;\;\;\;\;\;\;\;\;\;\;\;\;\text{{\rm s.t.}}\;\;\;
 \boldsymbol{g}_{-s,u,l}^{(s)} = \tilde{\boldsymbol{g}}_{-s,u,l}^{(j)}, \;\forall u,l, j \in \mathcal{G}_s \cup \{s\}, \label{consensus_constraint}\\
&\;\;\;\;\;\;\;\;\;\;\;\;\;\;\;\;\;\;\;\;\;\;\;\sum_{u = 1}^{U} \delta_{s,u}
\left\| \boldsymbol{w}_{s,u} \right\|^2  \le P_s, \label{per_pow_bud} \\
&\;\;\;\;\;\;\;\;\;\;\;\;\;\;\;\;\;\;\;\;\;\;\;\eqref{local_bf_constraint},
\end{align}
\end{subequations}
where $\Upsilon_{u}^{(s)} =
|1 - \mu_{u}^{(s)}\sum_{i=1}^{S}\bar{\alpha}_{i,u} g_{i,u,u}^{(s)}|^2 + |\mu_{u}^{(s)}|^2\Psi_u^{(s)}$.
Here, we have $\Psi_u^{(s)} =  \sum_{i=1}^{S}\beta_{i,u}|g_{i,u,u}^{(s)}|^2
+ \blue{\sum_{l\ne u}^{U}}\boldsymbol{g}_{u,l}^{(s)\mathsf{H}}\boldsymbol{T}_u\boldsymbol{g}_{u,l}^{(s)} + \sigma^2$.

The constraint in \eqref{consensus_constraint} enforces \emph{consensus} among the local copies of $g_{i,u,l}$ across all connected satellites, ensuring that each satellite's local copy, constrained by both its previous value and the neighbor-wise information, converges to a common value. Meanwhile, the constraint in \eqref{local_bf_constraint} ensures that $g_{s,u,l}^{(s)}$ is determined by the beamformers directly controlled by the $s$-th satellite and serves as a dummy variable.

It can be observed that \eqref{prob_refor_consensus} is not jointly convex with respect to all optimization variables. To handle this, we decompose the optimization process into two stages. In the \emph{outer iteration} stage, the auxiliary variables $\mu_u^{(s)}$ and $\nu_u^{(s)}$, introduced by the \ac{wmmse} framework, are updated. In the \emph{inner iteration} stage, these auxiliary variables are kept fixed while the satellites perform consensus updates over $\boldsymbol{g}_{-s,u,l}^{(s)}, \forall u,l$, using the \ac{cadmm} framework. The detailed procedures for the outer updates and the inner consensus updates carried out at the $s$-th satellite are described below.

\subsection{Outer Iteration via WMMSE}

\emph{1) Update of $\mu_{u}^{(s)}$:}  
Similar to \eqref{update_mu}, with all other variables fixed, the optimal $\mu_{u}^{(s)}$ is obtained as
\begin{equation}\label{update_mu_dist}
\mu_{u}^{(s)} = 
\frac{\left(\sum_{i=1}^{S}\bar{\alpha}_{i,u} g_{i,u,u}^{(s)}\right)^{\mathsf{*}}}
{\left|\sum_{i=1}^{S}\bar{\alpha}_{i,u} g_{i,u,u}^{(s)}\right|^2 + \Psi_u^{(s)}},\;\forall u.
\end{equation}

\emph{2) Update of $\nu_{u}^{(s)}$:}  
Analogous to \eqref{update_nu}, when all other variables are fixed, the optimal $\nu_{u}^{(s)}$ is given by
\begin{equation}\label{update_nu_dist}
\nu_{u}^{(s)} = \frac{1}{\Upsilon_{u}^{(s)}},\;\forall u.
\end{equation}

\begin{remark}
As will be elaborated later, under the \ac{cadmm} framework and for given $\mu_{u}^{(s)}$ and $\nu_{u}^{(s)}$, the exchange of $g_{i,u,l}^{(s)}$ among satellites enables information fusion across the network, ensuring that all local copies progressively converge to a common value, as long as the underlying topology, described by the graph $\mathcal{G}$, is connected. Once consensus is achieved, the superscript $(s)$ can be omitted from all variables, resulting in a fully decentralized and parallel solution to \eqref{prob_refor}. Furthermore, as indicated in \eqref{update_mu_dist} and \eqref{update_nu_dist}, the local auxiliaries $\mu_u^{(s)}$ and $\nu_u^{(s)}$ are determined entirely by $g_{i,u,l}^{(s)}$, so they align across satellites once the latter reach consensus, and no explicit consensus constraint is needed on them.
\end{remark}
 
\subsection{Inner Iteration via C-ADMM}

\subsubsection{Update of \texorpdfstring{$\boldsymbol{w}_{s,u}$}{w_u,l} and \texorpdfstring{$\boldsymbol{g}_{-s,u,l}^{(s)}$}{g_u,l}}

For fixed $\mu_u^{(s)}$ and $\nu_u^{(s)}$ in the outer iteration, the \emph{decentralized} inner iteration follows the \ac{cadmm} framework and is decomposed into \emph{local optimization} at each satellite and \emph{information exchange} among neighboring satellites in the \ac{leo} network. The consensus constraint \eqref{consensus_constraint} is incorporated into the objective via the local augmented Lagrangian at the $s$-th satellite, which reformulates \eqref{prob_refor_consensus} as
\begin{subequations}\label{prob_refor_consensus_Lag} 
\begin{align} 
&\mathop {\min }\limits_{\substack{\{\boldsymbol{w}_{s,u}\}_{\forall u \in \mathcal{U}_s},\\\{g_{i,u,l}^{(s)}\}_{\forall i \ne s,u,l \in \mathcal{U}_i}}} \sum_{u=1}^{U} \nu_{u}^{(s)} \Upsilon_{u}^{(s)}+\sum_{u,l=1}^{U} \sum_{j \in \mathcal{G}_s \cup \{s\}}\left(\Re \left\{\boldsymbol{z}_{-s,u,l}^{(j)\mathsf{H}} \right.\right. \notag \\ 
&\; \left.\left. \times \left(\boldsymbol{g}_{-s,u,l}^{(s)} - \tilde{\boldsymbol{g}}_{-s,u,l}^{(j)}\right)\right\} + \frac{\rho_{g}}{2}\left\|\boldsymbol{g}_{-s,u,l}^{(s)} - \tilde{\boldsymbol{g}}_{-s,u,l}^{(j)}\right\|_{\text{F}}^2\right)\\ 
&\;\;\;\;\;\;\;\;\;\;\;\;\;\;\;\; \text{{\rm s.t.}}\;\;\;\eqref{per_pow_bud},\eqref{local_bf_constraint}, 
\end{align} 
\end{subequations}
where $\boldsymbol{z}_{-s,u,l}^{(j)} = [z_{1,u,l}^{(j)},\ldots,z_{s-1,u,l}^{(j)},z_{s+1,u,l}^{(j)},\ldots,z_{S,u,l}^{(j)}]^{\mathsf{T}}$ denotes the local Lagrange multiplier at the $s$-th satellite, associated with enforcing consensus with $\tilde{\boldsymbol{g}}_{-s,u,l}^{(j)}$ in \eqref{consensus_constraint}, and $\rho_{g}$ is the corresponding penalty parameter.

Note that \eqref{prob_refor_consensus_Lag} is jointly convex in all involved optimization variables and thus constitutes a \ac{qcqp}, which can be directly solved using off-the-shelf solvers such as CVX.

\subsubsection{Update of Local Lagrange Multipliers \texorpdfstring{$\boldsymbol{z}_{-s,u,l}^{(j)}$}{z_u,l}}
The local Lagrange multipliers associated with the $j$-th consensus constraint, $\forall j \in \mathcal{G}_s \cup \{s\}$, are updated as

\begin{equation}\label{loc_lag_update}
\boldsymbol{z}_{-s,u,l}^{(j)} = \tilde{\boldsymbol{z}}_{-s,u,l}^{(j)} + \rho_g \left(\boldsymbol{g}_{-s,u,l}^{(s)} - \tilde{\boldsymbol{g}}_{-s,u,l}^{(j)}\right),\; \forall u,l, 
\end{equation}
where $ \tilde{\boldsymbol{z}}_{-s,u,l}^{(j)}$ represents the value of the local Lagrange multiplier from the previous iteration.

\subsubsection{Network-Wide Information Exchange} 
After completing the above local updates, each \ac{leo} satellite exchanges its local intermediate parameters, i.e., $\boldsymbol{g}_{u,l}^{(s)}$, with its neighboring satellites according to the \ac{isl} topology, and then proceeds to the next iteration. It is worth emphasizing that this iterative local information exchange over a connected network is essential for gradually diffusing information across the system and ultimately achieving network-wide consensus.

\subsection{Convergence, Signaling Overhead, and Complexity}\label{sec-overhead}
The overall decentralized procedure for solving \eqref{prob_ori} is summarized in Algorithm~\ref{wmmse_cadmm_cvx}.\footnote{Although the algorithm is presented in a canonical outer-inner structure for theoretical convergence, simulations show that running a single inner iteration per outer iteration preserves both convergence and ultimate performance, flattening the dual loop into a more implementation-friendly single loop.} Convergence follows from the well-established properties of the \ac{wmmse} framework and the \ac{cadmm}-based consensus updates~\cite{shi2011wmmse,gonzalo2010tsp}: for fixed $\mu_u$ and $\nu_u$, the \ac{cadmm} subroutine converges to a stationary point of \eqref{prob_refor_consensus}, since this subproblem is jointly convex in $\boldsymbol{w}_{s,u}$ and $\boldsymbol{g}_{-s,u,l}^{(s)}$~\cite{gonzalo2010tsp}, and the resulting sequence satisfies the conditions for convergence of the \ac{wmmse} procedure~\cite{shi2011wmmse}.

After each local update, every \ac{leo} satellite transmits its intermediate consensus variables $\boldsymbol{g}_{u,l}^{(s)}, \forall u,l$, to its neighbors. Nominally the per-satellite signaling overhead is $|\mathcal{G}_s|SU^2$, but $g_{i,u,l}^{(s)}=0$ whenever $l \notin \mathcal{U}_i$ allows transmitting only the non-zero entries, reducing it to $|\mathcal{G}_s||\mathcal{U}_s|SU$. In practice, $|\mathcal{U}_s|$ is typically bounded by the number of \acp{rfc}, which is much smaller than $U$. Moreover, since the entries of the intermediate variables are inner products between channels and beamformers, their dimension is independent of the antenna number $N$, which can be large in the massive-antenna regime discussed in Remark~\ref{cent_limit}. The signaling overhead is therefore kept at a modest and practically feasible level.

\begin{remark}\label{decent_cvx_complex}
The computational complexity of Algorithm~\ref{wmmse_cadmm_cvx} is dominated by solving \eqref{prob_refor_consensus_Lag}. Although this problem can be directly solved via CVX and each satellite only needs to optimize its own beamformers $\boldsymbol{w}_{s,u}, \forall u \in \mathcal{U}_s$, the introduction of the consensus auxiliary variables $g_{i,u,l}^{(s)}, \forall i\ne s, u, l\in\mathcal{U}_i$ substantially increases the dimensionality of the resulting optimization problem, yielding a per-iteration complexity of $\mathcal{O}((N|\mathcal{U}_s|+U\sum_{i\ne s}|\mathcal{U}_i|)^3)$. This complexity remains prohibitively high for local on-board updates, especially when many \acp{ut} are scheduled and many \ac{leo} satellites cooperate, and motivates the low-complexity solver developed next to ensure the scalability of the decentralized framework.
\end{remark}

\renewcommand{\algorithmicrequire}{\textbf{Input:}}
\renewcommand{\algorithmicensure}{\textbf{Output:}}
\begin{algorithm}[t]
\caption{Decentralized Cooperative Beamforming Design for Networked \ac{leo} Satellites}
\label{wmmse_cadmm_cvx}
\begin{algorithmic}[1]
\State \textbf{Initialize}: {$\boldsymbol{W}_s, \forall s$, 
$\boldsymbol{\delta}_s,\forall s$, $\boldsymbol{g}_{u,l}^{(s)}, \forall s, u, l$;}
\For {$s = 1 : S$ (\emph{in parallel})}
\Repeat {\emph{ Outer iteration via WMMSE}}
\State {Update $\mu_{u}^{(s)}$ using \eqref{update_mu_dist};}
\State {Update $\nu_{u}^{(s)}$ using \eqref{update_nu_dist};}
\Repeat {\emph{ Inner iteration via C-ADMM}}
\State {Update $\boldsymbol{w}_{s,u}$ and $\boldsymbol{g}_{u,l}^{(s)}$ by solving \eqref{prob_refor_consensus_Lag};}
\State{Update $\boldsymbol{z}_{-s,u,l}^{(j)}$ using \eqref{loc_lag_update};}
\State{Exchange $\boldsymbol{g}_{u,l}^{(s)}$ with neighbors via \acp{isl};}
\Until{the predefined convergence condition is met;}
\Until {the relative reduction in the objective value falls below a predefined threshold or a maximum number of iterations is reached;}
\EndFor
\State \textbf{Output}: {$\boldsymbol{w}_{s,u}, \forall s, u\in \mathcal{U}_s$.}
\end{algorithmic}
\end{algorithm}

\section{Low-Complexity Decentralized Solution}\label{sec_lc_scheme}
In this section, we develop a low-complexity solution to solve \eqref{prob_refor_consensus_Lag} optimally, thereby overcoming the main computational bottleneck in the overall decentralized design. The key idea is to transform \eqref{prob_refor_consensus_Lag} into an equivalent problem that \emph{depends only on the beamformers $\boldsymbol{w}_{s,u}$} by exploiting the optimal expression of $g_{i,u,l}^{(s)}$ as a function of $\boldsymbol{w}_{s,u}$. By leveraging the resulting problem structure, the transformed optimization can be efficiently solved via a low-complexity line search rather than relying on a generic solver.

\subsection{The Optimal Expression of \texorpdfstring{$\boldsymbol{g}_{-s,u,l}^{(s)}$}{g_{-s,u,l}^{(s)}}}
Let $\left[\boldsymbol{T}_u\right]_{-s,-s}$ and $\left[\boldsymbol{T}_u\right]_{-s,s}$ denote the matrices obtained by removing the $s$-th row and column of $\boldsymbol{T}_u$, and the vector obtained by removing the $s$-th entry of its $s$-th column, respectively. Moreover, let $\bar{\boldsymbol{\alpha}}_{-s,u} = [\bar{\alpha}_{1,u},\ldots,\bar{\alpha}_{s-1,u},\bar{\alpha}_{s+1,u},\ldots,\bar{\alpha}_{S,u}]^{\mathsf{T}}$ and  
$\boldsymbol{\beta}_{-s,u} = [\beta_{1,u},\ldots,\beta_{s-1,u},\beta_{s+1,u},\ldots,\beta_{S,u}]^{\mathsf{T}}$. We now present the following theorem, which characterizes the optimal expression of $\boldsymbol{g}_{-s,u,l}^{(s)}$ as a function of $\boldsymbol{w}_{s,l}$.

\begin{theorem}
For fixed $\boldsymbol{w}_{s,l}$, the optimal $\boldsymbol{g}_{-s,u,l}^{(s)}$ is given by
\begin{equation}\label{optimal_g_minus}
\boldsymbol{g}_{-s,u,l}^{(s)}=  \boldsymbol{\Gamma}_{u,l}^{(s)}\boldsymbol{w}_{s,l} + \boldsymbol{\zeta}_{u,l}^{(s)},\quad \forall u,l,
\end{equation}
where
\begin{align}\label{Xi_express} 
\boldsymbol{\Gamma}_{u,l}^{(s)}=\left\{ 
\begin{aligned} 
&-\nu_u^{(s)}\left|\mu_u^{(s)}\right|^2 {\boldsymbol{Q}_{u}^{(s)}}^{-1} \bar{\alpha}_{s,u}\bar{\boldsymbol{\alpha}}_{-s,u}^* \boldsymbol{b}^{\mathsf{T}}\left(\boldsymbol{\theta}_{s,u}\right) \delta_{s,l},\\
&\;\;\;\;\;\;\;\;\;\;\;\;\;\;\;\;\;\;\;\;\;\;\;\;\;\;\;\;\;\;\;\;\;\;\;\;\;\;\;\;\;\;\;\;\;\;\;\;\;\;\;\;\;\;\;\;\;\;\;\;\;\;\;\;\;\; l = u,\\ 
&-\nu_u^{(s)}\left|\mu_u^{(s)}\right|^2 {\boldsymbol{Q}_{u}^{(s)}}^{-1} \left[\boldsymbol{T}_u\right]_{-s,s} \boldsymbol{b}^{\mathsf{T}}\left(\boldsymbol{\theta}_{s,u}\right) \delta_{s,l}, \\
&\;\;\;\;\;\;\;\;\;\;\;\;\;\;\;\;\;\;\;\;\;\;\;\;\;\;\;\;\;\;\;\;\;\;\;\;\;\;\;\;\;\;\;\;\;\;\;\;\;\;\;\;\;\;\;\;\;\;\;\;\;\;\;\;\;\; l \ne u, 
\end{aligned} \right.\notag 
\end{align}
and 
\begin{align}
\boldsymbol{\zeta}_{u,l}^{(s)}=\left\{ 
\begin{aligned} 
&{\boldsymbol{Q}_{u}^{(s)}}^{-1}\left(\nu_u^{(s)}\mu_u^{(s)*}\bar{\boldsymbol{\alpha}}_{-s,u}^* +\frac{\rho_g}{2}\bar{\boldsymbol{g}}_{-s,u,l}^{(s)}  \right),& l = u,\\ 
&{\boldsymbol{Q}_{u}^{(s)}}^{-1}\frac{\rho_g}{2}\bar{\boldsymbol{g}}_{-s,u,l}^{(s)}, & l \ne u, 
\end{aligned} \right.\notag 
\end{align}
Here, we have
\begin{equation}
\bar{\boldsymbol{g}}_{-s,u,l}^{(s)} = \sum_{j \in \mathcal{G}_s \cup \{s\}} \left(\boldsymbol{g}_{-s,u,l}^{(j)}
- \frac{\boldsymbol{z}_{-s,u,l}^{(j)}}{\rho_g}\right) \notag   
\end{equation}
and
\begin{align}\label{Q_express} 
\boldsymbol{Q}_{u}^{(s)}=\left\{ 
\begin{aligned} 
&\nu_u^{(s)}\left|\mu_u^{(s)}\right|^2 \left(\bar{\boldsymbol{\alpha}}_{-s,u}^*\bar{\boldsymbol{\alpha}}_{-s,u}^{\mathsf{T}} + \mathrm{diag}\left(\boldsymbol{\beta}_{-s,u}\right)\right) ,\\
&+ \frac{\rho_g \left(\left|\mathcal{G}_s\right| + 1\right)\boldsymbol{I}_{S-1}}{2},& l = u,\\ 
&\nu_u^{(s)}\left|\mu_u^{(s)}\right|^2 \left[\boldsymbol{T}_u\right]_{-s,-s} + \frac{\rho_g  \left(\left|\mathcal{G}_s\right| + 1\right)\boldsymbol{I}_{S-1}}{2}, & l \ne u. 
\end{aligned} \right.\notag 
\end{align}

\end{theorem}

\emph{Proof Sketch:} By fixing $\boldsymbol{w}_{s,u}$ in \eqref{prob_refor_consensus_Lag}, we obtain the subproblem with respect to $\boldsymbol{g}_{-s,u,l}^{(s)}$ as
\begin{subequations}\label{subprob_fix_bf} 
\begin{align} 
&\mathop {\min }\limits_{\{g_{i,u,l}^{(s)}\}_{\forall i \ne s,u,l \in \mathcal{U}_i}} \sum_{u=1}^{U} \nu_{u}^{(s)} \Upsilon_{u}^{(s)}+\sum_{u,l=1}^{U} \sum_{j \in \mathcal{G}_s \cup \{s\}}\left(\Re \left\{\boldsymbol{z}_{-s,u,l}^{(j)\mathsf{H}} \right.\right. \notag \\ 
&\; \left.\left. \times \left(\boldsymbol{g}_{-s,u,l}^{(s)} - \tilde{\boldsymbol{g}}_{-s,u,l}^{(j)}\right)\right\} + \frac{\rho_{g}}{2}\left\|\boldsymbol{g}_{-s,u,l}^{(s)} - \tilde{\boldsymbol{g}}_{-s,u,l}^{(j)}\right\|_{\text{F}}^2\right)\\ 
&\;\;\;\;\;\;\;\;\;\;\;\;\;\;\;\; \text{{\rm s.t.}}\;\;\;\eqref{local_bf_constraint}.
\end{align}     
\end{subequations}

Through a series of algebraic manipulations, \eqref{subprob_fix_bf} can be decomposed into $U^2$ independent quadratic optimization subproblems as
\begin{equation}\label{separate_g}
\mathop{\min}_{\boldsymbol{g}_{-s,u,l}^{(s)}} 
\;\boldsymbol{g}_{-s,u,l}^{(s)\mathsf{H}} \boldsymbol{Q}_{u}^{(s)} \boldsymbol{g}_{-s,u,l}^{(s)}
- 2 \Re\!\left\{\boldsymbol{f}_{-s,u,l}^{(s)\mathsf{H}} \boldsymbol{g}_{-s,u,l}^{(s)}\right\},\; \forall u,l,
\end{equation}
where
\begin{align}\label{b_expression} 
\boldsymbol{f}_{-s,u,l}^{(s)}\!=\!\left\{ 
\begin{aligned} 
&\nu_u^{(s)} \left(\mu_u^{(s)*}-\left|\mu_u^{(s)}\right|^2 \bar{\alpha}_{s,u} g_{s,u,l}^{(s)}\right)\bar{\boldsymbol{\alpha}}_{-s,u}^* \\ 
&+ \frac{\rho_g}{2}\bar{\boldsymbol{g}}_{-s,u,l}^{(s)},& l = u,\\ 
&- \nu_u^{(s)}\left|\mu_u^{(s)}\right|^2 g_{s,u,l}^{(s)}\left[\boldsymbol{T}_u\right]_{-s,s}+\frac{\rho_g}{2}\bar{\boldsymbol{g}}_{-s,u,l}^{(s)} , & l \ne u. \end{aligned} \right.\notag 
\end{align}

It is straightforward to verify that $\boldsymbol{Q}_{u}^{(s)}, \forall u,l$, are positive semi-definite matrices, which directly leads to the closed-form optimal solution of \eqref{separate_g} as
\begin{equation}\label{optimal_g_inter}
\boldsymbol{g}_{-s,u,l}^{(s)}={\boldsymbol{Q}_{u}^{(s)}}^{-1} \boldsymbol{f}_{-s,u,l}^{(s)}.
\end{equation}
By substituting the expressions of $\boldsymbol{f}_{-s,u,l}^{(s)}, \forall u,l,$ into \eqref{optimal_g_inter} and invoking \eqref{local_bf_constraint}, we obtain \eqref{optimal_g_minus}.
\hfill $\blacksquare$

\subsection{Reformulation of \eqref{prob_refor_consensus_Lag} in Terms of \texorpdfstring{$\boldsymbol{w}_{s,u}$}{w_s,u} Only}
With the optimal expression of $\boldsymbol{g}_{-s,u,l}^{(s)}$ as a function of \blue{$\boldsymbol{w}_{s,l}$} given in \eqref{optimal_g_minus}, we can eliminate $\boldsymbol{g}_{-s,u,l}^{(s)}$ from \eqref{prob_refor_consensus_Lag}. 
Let $\bar{\boldsymbol{\alpha}}_{u} = [\bar{\alpha}_{1,u},\ldots,\bar{\alpha}_{S,u}]^{\mathsf{T}}$ and  
$\boldsymbol{\beta}_{u} = [\beta_{1,u},\ldots,\beta_{S,u}]^{\mathsf{T}}$. The resulting reduced formulation is summarized in the following theorem:

\begin{theorem}
Problem \eqref{prob_refor_consensus_Lag} is equivalent to 
\begin{subequations}\label{equi_prob_w} 
\begin{align} 
\mathop{\min}_{\{\boldsymbol{w}_{s,l}\}_{\forall l \in \mathcal{U}_s}} 
\;& \sum_{l=1}^{U} \left(\boldsymbol{w}_{s,l}^{\mathsf{H}} \boldsymbol{\Theta}_{s,l} \boldsymbol{w}_{s,l}
- 2 \Re \left\{\boldsymbol{\xi}_{s,l}^{\mathsf{H}} \boldsymbol{w}_{s,l}\right\}\right)\\ 
\text{{\rm s.t.}}\;\;\;& \eqref{per_pow_bud},
\end{align}     
\end{subequations}
where
\begin{align}
\boldsymbol{\Theta}_{s,l} = \sum_{u=1}^{U}&\left[\frac{\rho_g\left(\left|\mathcal{G}_s\right| + 1\right)}{2}\boldsymbol{\Gamma}_{u,l}^{(s)\mathsf{H}}\boldsymbol{\Gamma}_{u,l}^{(s)} \right.\notag\\
&\left. + \nu_u^{(s)}\left|\mu_u^{(s)}\right|^2 \boldsymbol{\Omega}_{u,l}^{(s)\mathsf{H}} \boldsymbol{X}_{u,l} \boldsymbol{\Omega}_{u,l}^{(s)}\right], \notag
\end{align}
with
\begin{equation}
\boldsymbol{X}_{u,l} =
\begin{cases}
\bar{\boldsymbol{\alpha}}_{u}^*\bar{\boldsymbol{\alpha}}_{u}^{\mathsf{T}} + \mathrm{diag}\left(\boldsymbol{\beta}_{u}\right), & u = l, \\[1mm]
\boldsymbol{T}_u, & u \ne l,
\end{cases}\notag
\end{equation}
and
\begin{align}
\boldsymbol{\xi}_{s,l} = \sum_{u=1}^{U}&\left[\frac{\rho_g\left(\left|\mathcal{G}_s\right| + 1\right)}{2}\boldsymbol{\Gamma}_{u,l}^{(s)\mathsf{H}}\!\left(\frac{\bar{\boldsymbol{g}}_{-s,u,l}^{(s)}}{\left|\mathcal{G}_s\right| + 1} - \boldsymbol{\zeta}_{u,l}^{(s)}\right) \right.\notag\\
&\left. + \boldsymbol{Y}_{u,l}^{(s)}\right], \notag
\end{align}
with
\begin{equation}
\boldsymbol{Y}_{u,l}^{(s)} =
\begin{cases}
\nu_u^{(s)}\!\left(\mu_u^{(s)*} - \left|\mu_u^{(s)}\right|^2 \bar{\boldsymbol{\alpha}}_{u}^{\mathsf{T}} \boldsymbol{\eta}_{u,l}^{(s)}\right) \boldsymbol{\Omega}_{u,l}^{(s)\mathsf{H}}\bar{\boldsymbol{\alpha}}_{u}^{*} \\
\quad - \nu_u^{(s)}\left|\mu_u^{(s)}\right|^2 \boldsymbol{\Omega}_{u,l}^{(s)\mathsf{H}} \mathrm{diag}\left(\boldsymbol{\beta}_u\right) \boldsymbol{\eta}_{u,l}^{(s)}, & u = l, \\[1mm]
-\nu_u^{(s)}\left|\mu_u^{(s)}\right|^2 \boldsymbol{\Omega}_{u,l}^{(s)\mathsf{H}} \boldsymbol{T}_u^{\mathsf{H}} \boldsymbol{\eta}_{u,l}^{(s)}, & u \ne l.
\end{cases}\notag
\end{equation}
Here, we have
\begin{align}\label{Omega_express} 
\boldsymbol{\Omega}_{u,l}^{(s)}=\left\{ 
\begin{aligned} 
&\left(\boldsymbol{e}_s-\nu_u^{(s)}\left|\mu_u^{(s)}\right|^2 \boldsymbol{E}_s{\boldsymbol{Q}_{u}^{(s)}}^{-1} \bar{\alpha}_{s,u}\bar{\boldsymbol{\alpha}}_{-s,u}^*\right)\\
&\times \boldsymbol{b}^{\mathsf{T}}\left(\boldsymbol{\theta}_{s,u}\right) \delta_{s,l},  \;\;\;\;\;\;\;\;\;\;\;\;\;\;\;\;\;\;\;\;\;\;\;\;\;\;\;\;\;\;\;\;\;\;\;\;\;\;\;\;\;\; l = u,\\ 
&\left(\boldsymbol{e}_s-\nu_u^{(s)}\left|\mu_u^{(s)}\right|^2 \boldsymbol{E}_s{\boldsymbol{Q}_{u}^{(s)}}^{-1} \left[\boldsymbol{T}_u\right]_{-s,s}\right)\\
&\times \boldsymbol{b}^{\mathsf{T}}\left(\boldsymbol{\theta}_{s,u}\right) \delta_{s,l}, \;\;\;\;\;\;\;\;\;\;\;\;\;\;\;\;\;\;\;\;\;\;\;\;\;\;\;\;\;\;\;\;\;\;\;\;\;\;\;\;\;\; l \ne u. 
\end{aligned} \right.\notag 
\end{align}
and 
\begin{equation}\label{Eta_express}
\boldsymbol{\eta}_{u,l}^{(s)}=
\begin{cases}
\boldsymbol{E}_s{\boldsymbol{Q}_{u}^{(s)}}^{-1}\left(\nu_u^{(s)}\mu_u^{(s)*}\bar{\boldsymbol{\alpha}}_{-s,u}^* +\frac{\rho_g}{2}\bar{\boldsymbol{g}}_{-s,u,l}^{(s)} \right), & l = u, \\[1mm]
\boldsymbol{E}_s{\boldsymbol{Q}_{u}^{(s)}}^{-1}\frac{\rho_g}{2}\bar{\boldsymbol{g}}_{-s,u,l}^{(s)}, & l \ne u,
\end{cases}\notag
\end{equation}
where $\boldsymbol{e}_s \in \mathbb{R}^S$ denotes the vector whose $s$-th entry is $1$ and all others are $0$, and  
$\boldsymbol{E}_s = [\boldsymbol{e}_1,\ldots,\boldsymbol{e}_{s-1},\boldsymbol{e}_{s+1},\ldots,\boldsymbol{e}_S]$. 
\end{theorem}

\emph{Proof Sketch:} Note that the full vector $\boldsymbol{g}_{u,l}^{(s)}$ can be written as
\begin{equation}\label{g_whole_vec}
\boldsymbol{g}_{u,l}^{(s)}
= \boldsymbol{E}_s \boldsymbol{g}_{-s,u,l}^{(s)}
+ g_{s,u,l}^{(s)} \boldsymbol{e}_s,
\quad \forall u,l.
\end{equation}
By substituting \eqref{optimal_g_minus} into \eqref{g_whole_vec}, we can readily obtain
\begin{equation}\label{optimal_g}
\boldsymbol{g}_{u,l}^{(s)}=  \boldsymbol{\Omega}_{u,l}^{(s)}\blue{\boldsymbol{w}_{s,l}} + \boldsymbol{\eta}_{u,l}^{(s)},\quad \forall u,l.
\end{equation}
Then, by substituting \eqref{optimal_g_minus} and \eqref{optimal_g} into \eqref{prob_refor_consensus_Lag} and performing a series of algebraic manipulations, we obtain \eqref{equi_prob_w}. \hfill $\blacksquare$

\begin{remark}
Note that $\boldsymbol{\Theta}_{s,l}, \forall l$, are positive semi-definite matrices. As a result, solving the \ac{qcqp} reformulation in \eqref{equi_prob_w}, which is free of $\boldsymbol{g}_{-s,u,l}^{(s)}$, incurs a dominant computational complexity of only $\mathcal{O}((N|\mathcal{U}_s|)^3)$, as opposed to $\mathcal{O}((N|\mathcal{U}_s| + (S-1)U^2)^3)$ required for solving the original problem in \eqref{prob_refor_consensus_Lag}. This represents a substantial reduction in the per-satellite computational burden, thereby significantly enhancing the algorithm's scalability. Moreover, by further exploiting the strong duality of this convex problem, we will show that an even lower-complexity solution can be obtained.
\end{remark}

\renewcommand{\algorithmicrequire}{\textbf{Input:}}
\renewcommand{\algorithmicensure}{\textbf{Output:}}
\begin{algorithm}[t]
\caption{Proposed Low-Complexity Solution to \eqref{prob_refor_consensus_Lag}}
\label{lc_solution}
\begin{algorithmic}[1]
\State \textbf{Initialize}: {$\mu_{u}^{(s)},\forall u$, $\nu_{u}^{(s)},\forall u$;}
\State {Obtain $\boldsymbol{\Theta}_{s}$ and $\boldsymbol{\xi}_{s}$;}
\State {Obtain eigenvalue decomposition of $\boldsymbol{\Theta}_{s,u},\forall u \in \mathcal{U}_s$;}
\State {Obtain $\lambda$ by linearly searching the zero point of \eqref{final_Search_form};}
\State {Obtain $\boldsymbol{w}_{s,u},\forall u \in \mathcal{U}_s$ via \eqref{optimal_bf_form_lc};}
\State \textbf{Output}: {$\boldsymbol{w}_{s,u},\forall u \in \mathcal{U}_s$, $\boldsymbol{g}_{u,l}^{(s)}, \forall u,l$.}
\end{algorithmic}
\end{algorithm}

\subsection{Quasi Closed-Form Solution}
Note that only the beamformers associated with the scheduled \acp{ut}, i.e., $\boldsymbol{w}_{s,l}, \forall l \in \mathcal{U}_s$, are involved in \eqref{equi_prob_w}. This is inherently guaranteed by the facts that $\boldsymbol{\Theta}_{s,l} = \boldsymbol{0}_{N\times N}, \forall l \notin \mathcal{U}_s$ and $\boldsymbol{\xi}_{s,l} = \boldsymbol{0}_{N}, \forall l \notin \mathcal{U}_s$. These properties follow from that $\boldsymbol{\Gamma}_{u,l}^{(s)} = \boldsymbol{0}_{(S-1)\times N}, \forall l \notin \mathcal{U}_s$ and $\boldsymbol{\Omega}_{u,l}^{(s)} = \boldsymbol{0}_{S \times N}, \forall l \notin \mathcal{U}_s$, since both terms contain the multiplier $\delta_{s,l}$, which is nonzero only when $l \in \mathcal{U}_s$. 

Let $u^{(s)}_1,\ldots,u^{(s)}_{|\mathcal{U}_s|}$ denote the indices of the \acp{ut} in $\mathcal{U}_s$. Then, denote $\boldsymbol{w}_s = [\boldsymbol{w}_{s,u^{(s)}_1}^{\mathsf{T}},\ldots,\boldsymbol{w}_{s,u^{(s)}_{|\mathcal{U}_s|}}^{\mathsf{T}}]^{\mathsf{T}}$, $\boldsymbol{\xi}_s =[\boldsymbol{\xi}_{s,u^{(s)}_1}^{\mathsf{T}},\ldots,\boldsymbol{\xi}_{s,u^{(s)}_{|\mathcal{U}_s|}}^{\mathsf{T}}]^{\mathsf{T}}$, and $\boldsymbol{\Theta}_s = \mathrm{diag}(\boldsymbol{\Theta}_{s,u^{(s)}_1},\ldots,\boldsymbol{\Theta}_{s,u^{(s)}_{|\mathcal{U}_s|}})$. 
We can recast \eqref{equi_prob_w} into the following more compact form as
\begin{subequations}\label{equi_prob_w_recast} 
\begin{align} 
\mathop{\min}_{\boldsymbol{w}_s} 
\;\;\;&  \boldsymbol{w}_s^{\mathsf{H}} \boldsymbol{\Theta}_{s} \boldsymbol{w}_s
- 2 \Re \left\{\boldsymbol{\xi}_{s}^{\mathsf{H}} \boldsymbol{w}_s\right\}\\ 
\text{{\rm s.t.}}\;\;\;& \left\|\boldsymbol{w}_s\right\|^2 \le P_s.
\end{align}     
\end{subequations} 

The Lagrangian of \eqref{equi_prob_w_recast} is formulated as
\begin{equation}
\mathcal{L}\left(\boldsymbol{w}_s, \lambda\right) =    \boldsymbol{w}_s^{\mathsf{H}}\boldsymbol{\Theta}_s\boldsymbol{w}_s  -2\Re\left\{\boldsymbol{\xi}_s^{\mathsf{H}} \boldsymbol{w}_{s}\right\} - \lambda \left(\left\|\boldsymbol{w}_s\right\|^2 - P_s\right), \notag
\end{equation}
where $\lambda \ge 0$ is the Lagrange multiplier associated with the power constraint. The optimal solution to \eqref{equi_prob_w_recast} can then be derived by examining its Karush-Kuhn-Tucker (KKT) conditions as
\begin{subequations}\label{kkt}
\begin{align}
\left(\boldsymbol{\Theta}_s + \lambda \boldsymbol{I}_{N|\mathcal{U}_s|} \right)\boldsymbol{w}_s &= \boldsymbol{\xi}_s, \label{kkt_main_R1}\\
\left\|\boldsymbol{w}_s\right\| &= \sqrt{P_s},\label{kkt_pow_R1}\\
\lambda &\ge 0.\label{kkt_lambda_R1}
\end{align}
\end{subequations}

It follows from \eqref{kkt} that
\begin{equation}\label{optimal_bf_form}
\boldsymbol{w}_s = \left(\boldsymbol{\Theta}_s + \lambda \boldsymbol{I}_{N|\mathcal{U}_s|}\right)^{-1} \boldsymbol{\xi}_s,    
\end{equation}
where $\lambda$ is chosen such that $\left\|(\boldsymbol{\Theta}_s + \lambda \boldsymbol{I}_{N|\mathcal{U}_s|})^{-1} \boldsymbol{\xi}_s\right\| = \sqrt{P_s}$. To this end, define the scalar function
\begin{equation}\label{search_func_R1}
h\left(\lambda\right) = \left\|\left(\boldsymbol{\Theta}_s + \lambda \boldsymbol{I}_{N|\mathcal{U}_s|}\right)^{-1} \boldsymbol{\xi}_s \right\|^2 - P_s.
\end{equation}
It is straightforward to verify that $h(\lambda)$ is monotonically decreasing with respect to $\lambda$ whenever $\boldsymbol{\Theta}_s \succeq \boldsymbol{0}_{N|\mathcal{U}_s|}$, which always holds. Consequently, a unique $\lambda$ can be efficiently determined using a simple line search method.

We note, however, that \eqref{search_func_R1} involves an $N|\mathcal{U}_s|$-dimensional matrix inversion, which incurs a computational complexity of $\mathcal{O}((N|\mathcal{U}_s|)^3)$ per iteration of the line search. To further reduce this complexity, we exploit the block-diagonal structure of $\boldsymbol{\Theta}_s$, which yields
\begin{align}
&\left(\underbrace{\mathrm{diag}\left(\boldsymbol{\Theta}_{s,u^{(s)}_1},\ldots,\boldsymbol{\Theta}_{s,u^{(s)}_{|\mathcal{U}_s|}}\right)}_{\boldsymbol{\Theta}_s}  + \lambda \boldsymbol{I}_{N|\mathcal{U}_s|}\right)^{-1}\\
&=\mathrm{diag}\left(\left(\boldsymbol{\Theta}_{s,u^{(s)}_1} + \lambda\boldsymbol{I}_{N}\right)^{-1},\ldots,\left(\boldsymbol{\Theta}_{s,u^{(s)}_{|\mathcal{U}_s|}}+ \lambda\boldsymbol{I}_{N}\right)^{-1}\right).    \notag
\end{align}

Denote the eigenvalue decomposition of $\boldsymbol{\Theta}_{s,u}$ as $\boldsymbol{\Theta}_{s,u} = \boldsymbol{U}_{s,u} \boldsymbol{\Lambda}_{s,u} \boldsymbol{U}_{s,u}^{\mathsf{H}}$, where $\boldsymbol{U}_{s,u} \in \mathbb{C}^{N \times N}$ is unitary and $\boldsymbol{\Lambda}_{s,u} = \mathrm{diag}[\omega_{s,u,1}, \ldots, \omega_{s,u,N}]$ contains the eigenvalues. Then, $h(\lambda)$ can be rewritten as
\begin{align}
h\left(\lambda\right) &= \left\|\left(\boldsymbol{\Theta}_s + \lambda \boldsymbol{I}_{N|\mathcal{U}_s|}\right)^{-1} \boldsymbol{\xi}_s \right\|^2 - P_s \notag \\
& = \sum_{u \in \mathcal{U}_s}\left\|\boldsymbol{U}_{s,u}\left(\boldsymbol{\Lambda}_{s,u} + \lambda \boldsymbol{I}_{N}\right)^{-1}\boldsymbol{U}_{s,u}^{\mathsf{H}}\boldsymbol{\xi}_{s,u}\right\|^2 - P_s \notag \\
&= \sum_{u \in \mathcal{U}_s}\sum_{n=1}^{N} \frac{\left|\varpi_{s,u,n}\right|^2}{\left(\omega_{s,u,n} + \lambda\right)^2}  - P_s, \label{final_Search_form}
\end{align}
where $\boldsymbol{\varpi}_{s,u} = [\varpi_{s,u,1},\ldots,\varpi_{s,u,N}]^\mathsf{T} = \boldsymbol{U}_{s,u}^{\mathsf{H}}\boldsymbol{\xi}_{s,u}$.

Finally, by exploiting the block-diagonal structure, \eqref{optimal_bf_form} reduces to the per-block form (for all $u \in \mathcal{U}_s$)
\begin{equation}\label{optimal_bf_form_lc}
\boldsymbol{w}_{s,u} = \boldsymbol{U}_{s,u}\mathrm{diag}\!\left(\frac{1}{\omega_{s,u,1}+\lambda},\ldots,\frac{1}{\omega_{s,u,N}+\lambda}\right)\boldsymbol{\varpi}_{s,u},
\end{equation}
which eliminates the need for an $N|\mathcal{U}_s|$-dimensional matrix inversion in both the line search and the final beamformer computation, thereby significantly reducing the overall computational complexity of the beamformer update.

\begin{figure}[t]
\centering
\includegraphics[width=1\linewidth]{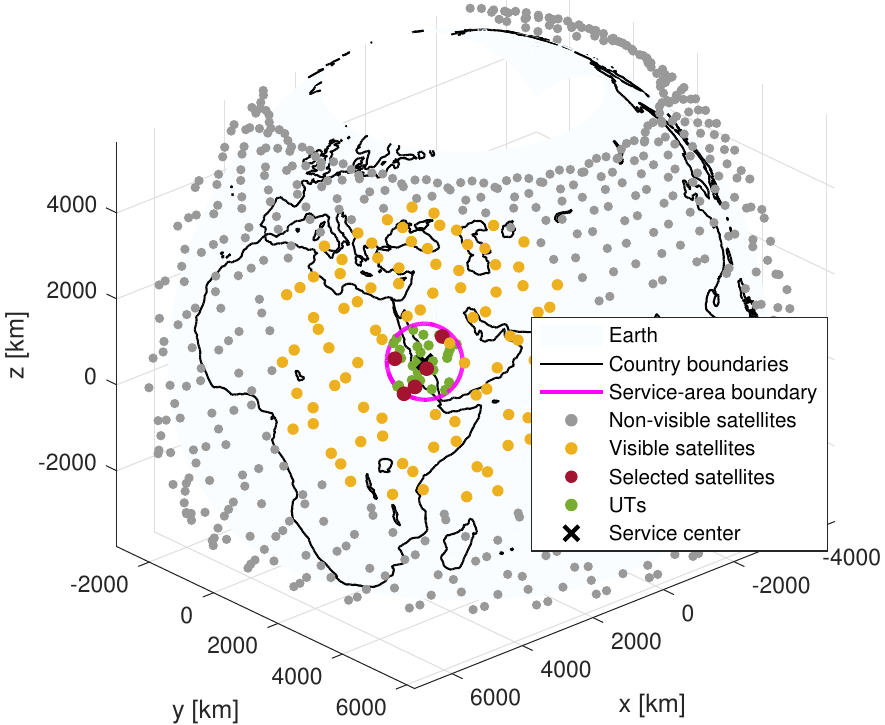}
\caption{An illustration of the system geometry.}\label{sys_geo}
\end{figure}

\subsection{Complexity}
The overall procedure for solving \eqref{prob_refor_consensus_Lag} using the proposed low-complexity solution is summarized in Algorithm~\ref{lc_solution}. The computational complexity is dominated by two main steps: (i) the pre-computation of $(\boldsymbol{Q}_{u}^{(s)})^{-1}, \forall u$, for constructing $\boldsymbol{\Theta}_{s}$ and $\boldsymbol{\xi}_{s}$, which incurs a complexity of $\mathcal{O}(U(S-1)^3)$; and (ii) the eigenvalue decomposition of $\boldsymbol{\Theta}_{s,u}, \forall u \in \mathcal{U}_s$, which incurs a complexity of $\mathcal{O}(N^3|\mathcal{U}_s|)$. In particular, when the number of antennas is significantly larger than the number of satellites, i.e., $N \gg S$, the cost of the first step becomes negligible. In contrast, solving \eqref{prob_refor_consensus_Lag} using CVX incurs a computational complexity on the order of $\mathcal{O}((N|\mathcal{U}_s| + U\sum_{i\ne s}|\mathcal{U}_i| )^3)$. This comparison clearly demonstrates the substantial superiority of the proposed solution in terms of reducing computational complexity and enhancing the scalability of decentralized networked \ac{leo} cooperative beamforming.

\section{Numerical Results}\label{sec_numer}

\subsection{Simulation Setting}
The Earth is modeled as a sphere with radius $6371\,\text{km}$. We consider \ac{leo} satellites operating at an orbital altitude of $550\,\text{km}$. As illustrated in Fig. \ref{sys_geo}, the satellite constellation follows a Walker-Delta configuration consisting of 28 orbit planes, each evenly populated with 60 satellites, and an orbital inclination of $53^{\circ}$\cite{yafei2026jsac}. We define a circular region of interest on the Earth's surface, centered at $20^{\circ}$ latitude and $40^{\circ}$ longitude with an $800\text{ km}$ radius, representing a high-demand service area in which the \acp{ut} are uniformly distributed. The cooperating set is taken to be the $S$ visible satellites nearest to the service-area centre at the considered snapshot. Each satellite is equipped with a \ac{upa} mounted tangentially to its orbital trajectory, with its local coordinate system oriented such that the boresight points toward the Earth's center. The large-scale path loss components $\gamma_{s,u}$ are generated according to the models specified in \cite{3gpp.38.811,zack2025twc}, while the Rician factors are randomly selected within the range of $15$ to $20\,\text{dB}$\cite{moewin2023jsac}. Unless otherwise stated, all remaining system and simulation parameters are provided in Table~\ref{simu_para}. The \ac{cadmm} penalty is initialised at $\rho_g=10^{3}$ and adapted multiplicatively up to $10^{5}$ whenever the consensus residual fails to decrease, following the standard adaptive heuristic\cite{gonzalo2010tsp}. Beamformers are initialised by per-satellite-power-normalized \ac{mrt}, the consensus variables to the corresponding consistent values, and the Lagrange multipliers to zero.

\begin{table}[t]
    \centering
    \caption{Simulation Parameters}
    \label{simu_para}
    \begin{tabular}{@{}ll@{}}
        \toprule
        \textbf{Parameter} & \textbf{Value} \\ 
        \midrule
        Carrier frequency $f_c$ & $5 \text{ GHz}$\cite{moewin2025jsac} \\
        Signal bandwidth $B$ & $20 \text{ MHz}$ \cite{moewin2025jsac,yafei2026jsac} \\
        Power budget at each \ac{leo} satellite & $50 \text{ dBm}$\\
        \ac{psd} $N_0$ & $-173.855 \text{ dBm/Hz}$ \\
        Noise figure $F$ & $10 \,\text{dB}$\\
        Number of \ac{leo} satellites $S$ & 5\\
        Number of \acp{ut} $U$ & 32\\
        Number of antennas $N = N_{\text{h}} \times N_{\text{v}}$ & $4 \times  4$\\
        Per-satellite maximum number of served \acp{ut} $U_{\rm{max}}$ & 8 \\    
        Antenna radiation gain (amplitude) $G(\theta)$ & $\sqrt{\frac{3}{2\pi}}\cos(\theta)$ \cite{balanis2005antenna} \\
        \bottomrule
    \end{tabular}
\end{table}

\subsection{Benchmark Schemes}

\subsubsection{Scheduling Schemes}
To assess the impact of user scheduling on the overall cooperative beamforming performance, we compare two pragmatic, geometry-driven scheduling strategies, in line with Remark~\ref{sched_decouple_remark}:
\begin{itemize}
    \item Correlation-based scheduling (CS): At each \ac{leo} satellite, the $U_{\rm{max}}$ \acp{ut} with the least channel correlation are selected. A simple greedy procedure is used: the closest \ac{ut} is first included, and additional users are added one by one by selecting the candidate with the minimum channel correlation to the current set, until $U_{\rm{max}}$ users are scheduled.
    \item Random scheduling (RS): At each \ac{leo} satellite, $U_{\rm{max}}$ users are selected uniformly at random from all $U$ \acp{ut}.
\end{itemize}

\subsubsection{Beamforming Schemes}
We evaluate the proposed decentralized scheme over three representative \ac{isl} topologies, Ring, Star, and Mesh, although the algorithm applies to \emph{any} connected \ac{isl} graph~\cite{liz2022access}. The centralized cooperative beamforming scheme developed in Section~\ref{sec_central} serves as a statistical-\ac{csi} upper bound. We further include the instantaneous \ac{csi}-based centralized scheme of~\cite{zack2026tcom} as an idealized upper bound. The gap between this curve and the statistical-\ac{csi} curves isolates the cost of operating without instantaneous \ac{csi}. For additional comparison, we consider two closed-form networked \ac{leo} cooperative beamforming baselines that do not require optimization:
\begin{itemize}
    \item \Ac{mrt}: At each \ac{leo} satellite, if $u \in \mathcal{U}_s$, the beamformer $\boldsymbol{w}_{s,u}$ is chosen to be parallel to $\boldsymbol{h}_{s,u}$; otherwise, $\boldsymbol{w}_{s,u} = \boldsymbol{0}_N$. A normalization factor is applied to satisfy the satellite power constraint.
    \item \Ac{zf}: At each \ac{leo} satellite, if $u \in \mathcal{U}_s$, the beamformer $\boldsymbol{w}_{s,u}$ is designed to lie in the null space of $\{\boldsymbol{h}_{s,j}\}_{j \notin \mathcal{U}_s}$, followed by a normalization step to meet the power constraint. Otherwise, $\boldsymbol{w}_{s,u} = \boldsymbol{0}_N$.
\end{itemize}

Additionally, to highlight the benefits of networked \ac{leo} cooperative beamforming, we also include a baseline based on \ac{sss}, in which each scheduled \ac{ut} is served by only a single satellite. To maintain consistency with the scheduling schemes described above, whenever a \ac{ut} is simultaneously selected by multiple satellites, only the strongest satellite-\ac{ut} link is retained. The beamformers at each satellite are then optimized independently, without any inter-satellite cooperation\cite{zack2025twc}.

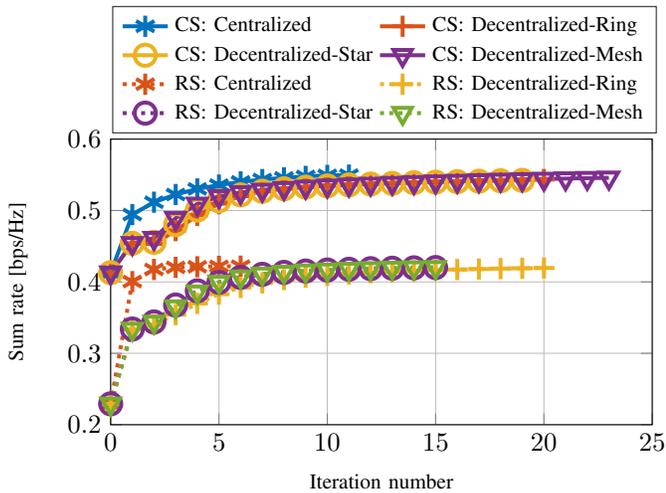
\begin{figure}[t]
\centering 
\centerline{
%
\definecolor{mycolor1}{rgb}{0.00000,0.44700,0.74100}%
\definecolor{mycolor2}{rgb}{0.85000,0.32500,0.09800}%
\definecolor{mycolor3}{rgb}{0.92900,0.69400,0.12500}%
\definecolor{mycolor4}{rgb}{0.49400,0.18400,0.55600}%
\definecolor{mycolor5}{rgb}{0.46600,0.67400,0.18800}%
\definecolor{mycolor6}{rgb}{0.30100,0.74500,0.93300}%
\definecolor{mycolor7}{rgb}{0.63500,0.07800,0.18400}%
\begin{tikzpicture}

\begin{axis}[%
width=72mm,
height=38mm,
at={(0mm, 0mm)},
scale only axis,
xmin=0,
xmax=17,
xlabel style={font=\color{white!15!black}, font=\footnotesize},
xlabel={Iteration number},
ymin=0.2,
ymax=0.8,
ylabel style={font=\color{white!15!black}, font=\footnotesize},
ylabel={Sum rate [bps/Hz]},
axis background/.style={fill=white},
xmajorgrids,
ymajorgrids,
legend style={at={(0.0029,1.02)}, anchor=south west, legend cell align=left, align=left, draw=white!15!black, font=\footnotesize, legend columns = 2}
]
\addplot [color=mycolor1, line width=1.5pt, mark size=4.0pt, mark=asterisk, mark options={solid, mycolor1}]
  table[row sep=crcr]{%
0	0.5715130968173203\\
1	0.7085406847328278\\
2	0.7327332376803462\\
3	0.7397739760458091\\
4	0.7424580504994891\\
5	0.7439514638167406\\
6	0.7451056549507719\\
7	0.7461612235368943\\
8	0.7471801406218176\\
9	0.7481659775104912\\
10	0.7491017518720885\\
11	0.749967545992092\\
12	0.750745948197402\\
13	0.7514243115307127\\
};
\addlegendentry{CS: Centralized}

\addplot [color=mycolor2, line width=1.5pt, mark size=4.0pt, mark=+, mark options={solid, mycolor2}]
  table[row sep=crcr]{%
0	0.5715130968173203\\
1	0.6546468174622779\\
2	0.6565657152286015\\
3	0.6949385365746082\\
4	0.717331929607474\\
5	0.7289071400268889\\
6	0.7351696862541159\\
7	0.7382539764586438\\
8	0.7397380339385518\\
9	0.7405384004077211\\
10	0.741039411746754\\
};
\addlegendentry{CS: Decentralized-Ring}

\addplot [color=mycolor3, line width=1.5pt, mark size=4.0pt, mark=o, mark options={solid, mycolor3}]
  table[row sep=crcr]{%
0	0.5715130968173203\\
1	0.6547333396194817\\
2	0.6569759109022022\\
3	0.6916095099102118\\
4	0.7143653882362957\\
5	0.7264037776104408\\
6	0.7331110405063013\\
7	0.736807126648676\\
8	0.7389336513592905\\
9	0.7402510827339432\\
10	0.7412451123625234\\
11	0.741902865855863\\
};
\addlegendentry{CS: Decentralized-Star}

\addplot [color=mycolor4, line width=1.5pt, mark size=4.0pt, mark=triangle, mark options={solid, mycolor4}]
  table[row sep=crcr]{%
0	0.5715130968173203\\
1	0.6547230696369393\\
2	0.6600906364824966\\
3	0.7013439028514005\\
4	0.7266802016455328\\
5	0.7358870419471127\\
6	0.7387409928190357\\
7	0.7399026368223296\\
8	0.7406633975272912\\
9	0.7412470158821467\\
};
\addlegendentry{CS: Decentralized-Mesh}

\addplot [color=mycolor2, line width=1.5pt, mark size=4.0pt, mark=asterisk, mark options={solid, mycolor2}, dotted]
  table[row sep=crcr]{%
0	0.28564604184653486\\
1	0.45112544046881753\\
2	0.48388406704661263\\
3	0.4955162502416798\\
4	0.5015649356837297\\
5	0.5050186359255094\\
6	0.5070431486567776\\
7	0.5082620978149012\\
8	0.5090199285612906\\
9	0.5095101816781292\\
};
\addlegendentry{RS: Centralized}

\addplot [color=mycolor3, line width=1.5pt, mark size=4.0pt, mark=+, mark options={solid, mycolor3}, dotted]
  table[row sep=crcr]{%
0	0.28564604184653486\\
1	0.40623533041582327\\
2	0.42959844747652676\\
3	0.4439557960504037\\
4	0.4580529083178834\\
5	0.47241357221099406\\
6	0.4843377212027169\\
7	0.49221284009180843\\
8	0.4967514711901843\\
9	0.49941687876504587\\
10	0.5012273660960208\\
11	0.5027487906761627\\
12	0.5040250032954037\\
13	0.5051335453222138\\
14	0.5060368040943029\\
15	0.5067616462327598\\
16	0.5073374134392479\\
17	0.5078329382844516\\
};
\addlegendentry{RS: Decentralized-Ring}

\addplot [color=mycolor4, line width=1.5pt, mark size=4.0pt, mark=o, mark options={solid, mycolor4}, dotted]
  table[row sep=crcr]{%
0	0.28564604184653486\\
1	0.40629146259394394\\
2	0.42272136122935833\\
3	0.44395079580594354\\
4	0.46631394286996636\\
5	0.48213219393759615\\
6	0.491705054470425\\
7	0.49727782757119093\\
8	0.500659243587904\\
9	0.5028254202357212\\
10	0.5043651237730902\\
11	0.5055301879740508\\
12	0.5064029582170111\\
13	0.5071139758803689\\
14	0.5077285840134976\\
15	0.508198821984919\\
};
\addlegendentry{RS: Decentralized-Star}

\addplot [color=mycolor5, line width=1.5pt, mark size=4.0pt, mark=triangle, mark options={solid, mycolor5}, dotted]
  table[row sep=crcr]{%
0	0.28564604184653486\\
1	0.4063065925606365\\
2	0.42452042638472054\\
3	0.4450729885892518\\
4	0.4662988933818749\\
5	0.48081222110929345\\
6	0.48909157938280584\\
7	0.4940197259213046\\
8	0.49739579515925075\\
9	0.4999815473787016\\
10	0.5020347812294479\\
11	0.5036488598475477\\
12	0.5049191813962837\\
13	0.5059775272075631\\
14	0.5067777219367476\\
15	0.5074861607680842\\
16	0.5080399553437507\\
17	0.5085381687672652\\
};
\addlegendentry{RS: Decentralized-Mesh}
\end{axis}
\end{tikzpicture}%
}
\vspace{-0.2cm}
\caption{Convergent behavior of the proposed networked \ac{leo} satellite cooperative beamforming schemes.}
\label{converge}
\end{figure}

\begin{figure}[t]
\centering
\begin{minipage}[b]{0.98\linewidth}
\vspace{-0.3cm}
  \centering
%
\definecolor{mycolor1}{rgb}{0.00000,0.44700,0.74100}%
\definecolor{mycolor2}{rgb}{0.85000,0.32500,0.09800}%
\definecolor{mycolor3}{rgb}{0.92900,0.69400,0.12500}%
\definecolor{mycolor4}{rgb}{0.49400,0.18400,0.55600}%
\definecolor{mycolor5}{rgb}{0.46600,0.67400,0.18800}%
\definecolor{mycolor6}{rgb}{0.30100,0.74500,0.93300}%
\definecolor{mycolor7}{rgb}{0.63500,0.07800,0.18400}%
\definecolor{mycolor8}{rgb}{0.10000,0.10000,0.10000}%
\begin{tikzpicture}

\begin{axis}[%
width=72mm,
height=38mm,
at={(0mm, 0mm)},
scale only axis,
xmin=40,
xmax=60,
xlabel style={font=\color{white!15!black}, font=\footnotesize},
xlabel={Per-satellite power budget $P_s$ [dBm]},
ymin=0,
ymax=5.0,
ylabel style={font=\color{white!15!black}, font=\footnotesize},
ylabel={Sum rate [bps/Hz]},
axis background/.style={fill=white},
xmajorgrids,
ymajorgrids,
legend style={at={(0.0,0.15)}, font=\footnotesize, anchor=south west, legend cell align=left, align=left, draw=white!15!black}
]
\addplot [color=mycolor8, line width=1.5pt, mark size=4.0pt, mark=triangle, mark options={solid, mycolor8}, densely dashed]
  table[row sep=crcr]{%
40.0	0.08918268242089894\\
45.0	0.2703919112571155\\
50.0	0.7765262048447147\\
55.0	2.0230276895198207\\
60.0	4.577698444443346\\
};
\addlegendentry{Centralized (iCSI)}

\addplot [color=mycolor1, line width=1.5pt, mark size=4.0pt, mark=o, mark options={solid, mycolor1}]
  table[row sep=crcr]{%
40.0	0.0859339095408382\\
45.0	0.260311580984858\\
50.0	0.7514243115307127\\
55.0	1.977083833573908\\
60.0	4.519775951138131\\
};
\addlegendentry{Centralized (sCSI)}

\addplot [color=mycolor2, line width=1.5pt, mark size=4.0pt, mark=+, mark options={solid, mycolor2}]
  table[row sep=crcr]{%
40.0	0.08501464244258239\\
45.0	0.25807896543292397\\
50.0	0.741039411746754\\
55.0	1.9495626538839403\\
60.0	4.507403494661167\\
};
\addlegendentry{Decentralized-Ring}

\addplot [color=mycolor3, line width=1.5pt, mark size=4.0pt, mark=square, mark options={solid, mycolor3}]
  table[row sep=crcr]{%
40.0	0.08476274609193943\\
45.0	0.25779668100955255\\
50.0	0.741902865855863\\
55.0	1.9470825974162236\\
60.0	4.506990912075006\\
};
\addlegendentry{Decentralized-Star}

\addplot [color=mycolor4, line width=1.5pt, mark size=4.0pt, mark=diamond, mark options={solid, mycolor4}]
  table[row sep=crcr]{%
40.0	0.08504409871179752\\
45.0	0.25817146119163564\\
50.0	0.7412470158821467\\
55.0	1.9568694234262543\\
60.0	4.509959482025559\\
};
\addlegendentry{Decentralized-Mesh}

\addplot [color=mycolor5, line width=1.5pt, mark size=4.0pt, mark=asterisk, mark options={solid, mycolor5}]
  table[row sep=crcr]{%
40.0	0.06160034789680361\\
45.0	0.1911649491671072\\
50.0	0.5715130968173203\\
55.0	1.5532181865500263\\
60.0	3.508647277461286\\
};
\addlegendentry{MRT}

\addplot [color=mycolor6, line width=1.5pt, mark size=4.0pt, mark=x, mark options={solid, mycolor6}]
  table[row sep=crcr]{%
40.0	0.009369865861834685\\
45.0	0.029557735575749863\\
50.0	0.09275804866483839\\
55.0	0.28656733285690716\\
60.0	0.8479303866370403\\
};
\addlegendentry{ZF}

\addplot [color=mycolor7, line width=1.5pt, mark size=4.0pt, mark=star, mark options={solid, mycolor7}]
  table[row sep=crcr]{%
40.0	0.023781021468720513\\
45.0	0.07487211648955251\\
50.0	0.23354341419862695\\
55.0	0.7085375455412554\\
60.0	1.9952076116629238\\
};
\addlegendentry{SSS}
\end{axis}
\end{tikzpicture}%
    \vspace{-1.cm}
  \centerline{(a) Sum rate versus power budget $P_s$} \medskip
\end{minipage}
\hfill
\begin{minipage}[b]{0.98\linewidth}
  \centering
%
\definecolor{mycolor1}{rgb}{0.00000,0.44700,0.74100}%
\definecolor{mycolor2}{rgb}{0.85000,0.32500,0.09800}%
\definecolor{mycolor3}{rgb}{0.92900,0.69400,0.12500}%
\definecolor{mycolor4}{rgb}{0.49400,0.18400,0.55600}%
\definecolor{mycolor5}{rgb}{0.46600,0.67400,0.18800}%
\definecolor{mycolor6}{rgb}{0.30100,0.74500,0.93300}%
\definecolor{mycolor7}{rgb}{0.63500,0.07800,0.18400}%
\definecolor{mycolor8}{rgb}{0.10000,0.10000,0.10000}%
\begin{tikzpicture}

\begin{axis}[%
width=72mm,
height=38mm,
at={(0mm, 0mm)},
scale only axis,
xtick={9, 16, 25, 36, 49, 64},
xmin=9,
xmax=64,
xlabel style={font=\color{white!15!black}, font=\footnotesize},
xlabel={Per-satellite antenna number $N$},
ymin=0,
ymax=3.0,
ylabel style={font=\color{white!15!black}, font=\footnotesize},
ylabel={Sum rate [bps/Hz]},
axis background/.style={fill=white},
xmajorgrids,
ymajorgrids,
]
\addplot [color=mycolor8, line width=1.5pt, mark size=4.0pt, mark=triangle, mark options={solid, mycolor8}, densely dashed]
  table[row sep=crcr]{%
9	0.4592232121223643\\
16	0.7765260548920159\\
25	1.1509691973879888\\
36	1.5703319376706633\\
49	2.02554414889707\\
64	2.510109893671189\\
};

\addplot [color=mycolor1, line width=1.5pt, mark size=4.0pt, mark=o, mark options={solid, mycolor1}]
  table[row sep=crcr]{%
9	0.44210682372577065\\
16	0.7514243115307127\\
25	1.1156930140030181\\
36	1.5244271969607306\\
49	1.9667819556708164\\
64	2.4462677900386978\\
};

\addplot [color=mycolor2, line width=1.5pt, mark size=4.0pt, mark=+, mark options={solid, mycolor2}]
  table[row sep=crcr]{%
9	0.4369252754101018\\
16	0.741039411746754\\
25	1.101239266998689\\
36	1.5097003065714065\\
49	1.793803068594121\\
64	2.247277127877754\\
};

\addplot [color=mycolor3, line width=1.5pt, mark size=4.0pt, mark=square, mark options={solid, mycolor3}]
  table[row sep=crcr]{%
9	0.4368071144843065\\
16	0.741902865855863\\
25	1.1034830354913692\\
36	1.5106178284553367\\
49	1.9542551782643014\\
64	2.4289830239904684\\
};

\addplot [color=mycolor4, line width=1.5pt, mark size=4.0pt, mark=diamond, mark options={solid, mycolor4}]
  table[row sep=crcr]{%
9	0.43705231353519647\\
16	0.7412470158821467\\
25	1.1022556341466605\\
36	1.5101733091333105\\
49	1.9583084232167793\\
64	2.4408055409195466\\
};

\addplot [color=mycolor5, line width=1.5pt, mark size=4.0pt, mark=asterisk, mark options={solid, mycolor5}]
  table[row sep=crcr]{%
9	0.32681533394334783\\
16	0.5715130968173203\\
25	0.8734734630561928\\
36	1.2243312217284101\\
49	1.6190773271462806\\
64	2.0511460787465463\\
};

\addplot [color=mycolor6, line width=1.5pt, mark size=4.0pt, mark=x, mark options={solid, mycolor6}]
  table[row sep=crcr]{%
9	0.006852796871864644\\
16	0.09275804866483839\\
25	0.2445775454936971\\
36	0.4581783128728335\\
49	0.7424093890638365\\
64	1.0976595939081895\\
};

\addplot [color=mycolor7, line width=1.5pt, mark size=4.0pt, mark=star, mark options={solid, mycolor7}]
  table[row sep=crcr]{%
9	0.13221818547695707\\
16	0.23354341419862695\\
25	0.3616513925553383\\
36	0.5158969194980139\\
49	0.6952643839913677\\
64	0.8969436003614011\\
};
\end{axis}
\end{tikzpicture}%
    \vspace{-1.cm}
  \centerline{(b) Sum rate versus antenna number $N$} \medskip
\end{minipage}
\hfill
\begin{minipage}[b]{0.98\linewidth}
  \centering
%
\definecolor{mycolor1}{rgb}{0.00000,0.44700,0.74100}%
\definecolor{mycolor2}{rgb}{0.85000,0.32500,0.09800}%
\definecolor{mycolor3}{rgb}{0.92900,0.69400,0.12500}%
\definecolor{mycolor4}{rgb}{0.49400,0.18400,0.55600}%
\definecolor{mycolor5}{rgb}{0.46600,0.67400,0.18800}%
\definecolor{mycolor6}{rgb}{0.30100,0.74500,0.93300}%
\definecolor{mycolor7}{rgb}{0.63500,0.07800,0.18400}%
\definecolor{mycolor8}{rgb}{0.10000,0.10000,0.10000}%
\begin{tikzpicture}

\begin{axis}[%
width=72mm,
height=38mm,
at={(0mm, 0mm)},
scale only axis,
xmin=4,
xmax=8,
xlabel style={font=\color{white!15!black}, font=\footnotesize},
xlabel={LEO satellite number $S$},
ymin=0,
ymax=1.4000000000000001,
ylabel style={font=\color{white!15!black}, font=\footnotesize},
ylabel={Sum rate [bps/Hz]},
axis background/.style={fill=white},
xmajorgrids,
ymajorgrids,
]
\addplot [color=mycolor8, line width=1.5pt, mark size=4.0pt, mark=triangle, mark options={solid, mycolor8}, densely dashed]
  table[row sep=crcr]{%
4	0.56868133166286\\
5	0.7765262048447147\\
6	0.9612018313748333\\
7	1.1201899937802375\\
8	1.2471064024046932\\
};

\addplot [color=mycolor1, line width=1.5pt, mark size=4.0pt, mark=o, mark options={solid, mycolor1}]
  table[row sep=crcr]{%
4	0.5246172791005463\\
5	0.7514243115307127\\
6	0.9061201034494418\\
7	1.0696496907843185\\
8	1.2095270610086077\\
};

\addplot [color=mycolor2, line width=1.5pt, mark size=4.0pt, mark=+, mark options={solid, mycolor2}]
  table[row sep=crcr]{%
4	0.5164470118923847\\
5	0.741039411746754\\
6	0.9024251184026034\\
7	1.0656285887128845\\
8	1.1995860206485092\\
};

\addplot [color=mycolor3, line width=1.5pt, mark size=4.0pt, mark=square, mark options={solid, mycolor3}]
  table[row sep=crcr]{%
4	0.45669596587717387\\
5	0.741902865855863\\
6	0.9031486285468973\\
7	1.0649302223586379\\
8	1.195056478416626\\
};

\addplot [color=mycolor4, line width=1.5pt, mark size=4.0pt, mark=diamond, mark options={solid, mycolor4}]
  table[row sep=crcr]{%
4	0.5171871771388787\\
5	0.7412470158821467\\
6	0.9021398555233804\\
7	1.0653340770220463\\
8	1.2014418855340598\\
};

\addplot [color=mycolor5, line width=1.5pt, mark size=4.0pt, mark=asterisk, mark options={solid, mycolor5}]
  table[row sep=crcr]{%
4	0.41803309437963987\\
5	0.5715130968173203\\
6	0.7192735820808585\\
7	0.8461857551323505\\
8	1.0091525831711774\\
};

\addplot [color=mycolor6, line width=1.5pt, mark size=4.0pt, mark=x, mark options={solid, mycolor6}]
  table[row sep=crcr]{%
4	0.07550826423069221\\
5	0.09275804866483839\\
6	0.10359832522034444\\
7	0.10610122722348773\\
8	0.11826091519653957\\
};

\addplot [color=mycolor7, line width=1.5pt, mark size=4.0pt, mark=star, mark options={solid, mycolor7}]
  table[row sep=crcr]{%
4	0.18794610778440146\\
5	0.23354341419862695\\
6	0.23414548735904533\\
7	0.31158567977893864\\
8	0.35657942390826647\\
};
\end{axis}
\end{tikzpicture}%
    \vspace{-1.cm}
  \centerline{(c) Sum rate versus \ac{leo} satellite number $S$} \medskip
\end{minipage}
\vspace{-0.4cm}
\caption{Sum rate comparison under various schemes.}
\label{SR_comparison}
\vspace{-0.2cm}
\end{figure}

\subsection{Simulation Results}
\subsubsection{Convergence}
In Fig. \ref{converge}, we evaluate the convergence behavior of the proposed decentralized cooperative beamforming scheme under different \ac{isl} topologies and scheduling strategies. Across all considered settings, the decentralized schemes converge to sum-rate values that closely match the centralized benchmark, demonstrating their effectiveness in approaching the upper performance bound. Among the three \ac{isl} topologies, the performance gap remains modest, highlighting the robustness and versatility of the proposed framework. Regarding the impact of scheduling, the RS-based schemes unsurprisingly yield substantially lower performance than the CS-based ones, since CS reduces \ac{iui} by selecting users with lower spatial correlation, thereby enhancing the effectiveness of cooperative beamforming.

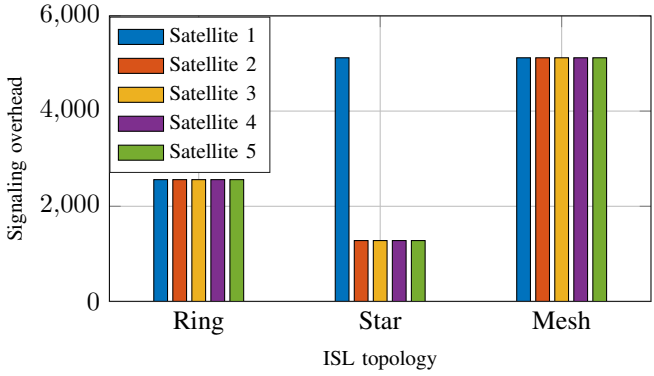
\begin{figure}[t]
\centering 
\centerline{
%
\definecolor{mycolor1}{rgb}{0.00000,0.44700,0.74100}%
\definecolor{mycolor2}{rgb}{0.85000,0.32500,0.09800}%
\definecolor{mycolor3}{rgb}{0.92900,0.69400,0.12500}%
\definecolor{mycolor4}{rgb}{0.49400,0.18400,0.55600}%
\definecolor{mycolor5}{rgb}{0.46600,0.67400,0.18800}%
\begin{tikzpicture}

\begin{axis}[%
width=72mm,
height=38mm,
at={(0mm, 0mm)},
scale only axis,
bar shift auto,
xmin=0.507692307692308,
xmax=3.49230769230769,
xtick={1,2,3},
xticklabels={{Ring},{Star},{Mesh}},
xlabel style={font=\color{white!15!black}, font=\footnotesize},
xlabel={ISL topology},
ymin=0,
ymax=6000,
ylabel style={font=\color{white!15!black}, font=\footnotesize},
ylabel={Signaling overhead},
axis background/.style={fill=white},
xmajorgrids,
ymajorgrids,
legend style={at={(0.0,0.450)}, font=\footnotesize, anchor=south west, legend cell align=left, align=left, draw=white!15!black}
]
\addplot[ybar, bar width=5.20, fill=mycolor1, draw=black, area legend] table[row sep=crcr] {%
1	2560\\
2	5120\\
3	5120\\
};
\addplot[forget plot, color=white!15!black] table[row sep=crcr] {%
0.507692307692308	0\\
3.49230769230769	0\\
};
\addlegendentry{Satellite 1}

\addplot[ybar, bar width=5.20, fill=mycolor2, draw=black, area legend] table[row sep=crcr] {%
1	2560\\
2	1280\\
3	5120\\
};
\addplot[forget plot, color=white!15!black] table[row sep=crcr] {%
0.507692307692308	0\\
3.49230769230769	0\\
};
\addlegendentry{Satellite 2}

\addplot[ybar, bar width=5.20, fill=mycolor3, draw=black, area legend] table[row sep=crcr] {%
1	2560\\
2	1280\\
3	5120\\
};
\addplot[forget plot, color=white!15!black] table[row sep=crcr] {%
0.507692307692308	0\\
3.49230769230769	0\\
};
\addlegendentry{Satellite 3}

\addplot[ybar, bar width=5.20, fill=mycolor4, draw=black, area legend] table[row sep=crcr] {%
1	2560\\
2	1280\\
3	5120\\
};
\addplot[forget plot, color=white!15!black] table[row sep=crcr] {%
0.507692307692308	0\\
3.49230769230769	0\\
};
\addlegendentry{Satellite 4}

\addplot[ybar, bar width=5.20, fill=mycolor5, draw=black, area legend] table[row sep=crcr] {%
1	2560\\
2	1280\\
3	5120\\
};
\addplot[forget plot, color=white!15!black] table[row sep=crcr] {%
0.507692307692308	0\\
3.49230769230769	0\\
};
\addlegendentry{Satellite 5}
\end{axis}
\end{tikzpicture}%
}
\vspace{-0.2cm}
\caption{Per-satellite signaling overhead versus \ac{isl} topologies.}
\label{Overhead_topo}
\end{figure}

\begin{figure}[t]
\centering
\begin{minipage}[b]{0.98\linewidth}
\vspace{-0.3cm}
  \centering
%
\definecolor{mycolor1}{rgb}{0.00000,0.44700,0.74100}%
\definecolor{mycolor2}{rgb}{0.85000,0.32500,0.09800}%
\definecolor{mycolor3}{rgb}{0.92900,0.69400,0.12500}%
\definecolor{mycolor4}{rgb}{0.49400,0.18400,0.55600}%
\definecolor{mycolor5}{rgb}{0.46600,0.67400,0.18800}%
\definecolor{mycolor6}{rgb}{0.30100,0.74500,0.93300}%
\definecolor{mycolor7}{rgb}{0.63500,0.07800,0.18400}%
\begin{tikzpicture}

\begin{axis}[%
width=72mm,
height=38mm,
at={(0mm, 0mm)},
scale only axis,
xtick={4, 8, 12, 16, 20, 24, 28, 32},
xmin=4,
xmax=32,
xlabel style={font=\color{white!15!black}, font=\footnotesize},
xlabel={UT number $U$},
ymin=0,
ymax=6000,
ylabel style={font=\color{white!15!black}, font=\footnotesize},
ylabel={Signaling overhead},
axis background/.style={fill=white},
xmajorgrids,
ymajorgrids,
yminorgrids,
legend style={at={(0.0,0.66)}, font=\footnotesize, anchor=south west, legend cell align=left, align=left, draw=white!15!black, legend columns = 3}
]
\addplot [color=mycolor1, line width=1.5pt, mark size=4.0pt, mark=o, mark options={solid, mycolor1}]
  table[row sep=crcr]{%
4	160.0\\
8	640.0\\
12	960.0\\
16	1280.0\\
20	1600.0\\
24	1920.0\\
28	2240.0\\
32	2560.0\\
};
\addlegendentry{Ring (mean)}

\addplot [color=mycolor2, line width=1.5pt, mark size=4.0pt, mark=o, mark options={solid, mycolor2}, dashed]
  table[row sep=crcr]{%
4	160\\
8	640\\
12	960\\
16	1280\\
20	1600\\
24	1920\\
28	2240\\
32	2560\\
};
\addlegendentry{Ring (max)}

\addplot [color=mycolor3, line width=1.5pt, mark size=4.0pt, mark=o, mark options={solid, mycolor3}, dotted]
  table[row sep=crcr]{%
4	160\\
8	640\\
12	960\\
16	1280\\
20	1600\\
24	1920\\
28	2240\\
32	2560\\
};
\addlegendentry{Ring (min)}

\addplot [color=mycolor4, line width=1.5pt, mark size=4.0pt, mark=square, mark options={solid, mycolor4}]
  table[row sep=crcr]{%
4	128.0\\
8	512.0\\
12	768.0\\
16	1024.0\\
20	1280.0\\
24	1536.0\\
28	1792.0\\
32	2048.0\\
};
\addlegendentry{Star (mean)}

\addplot [color=mycolor5, line width=1.5pt, mark size=4.0pt, mark=square, mark options={solid, mycolor5}, dashed]
  table[row sep=crcr]{%
4	320\\
8	1280\\
12	1920\\
16	2560\\
20	3200\\
24	3840\\
28	4480\\
32	5120\\
};
\addlegendentry{Star (max)}

\addplot [color=mycolor6, line width=1.5pt, mark size=4.0pt, mark=square, mark options={solid, mycolor6}, dotted]
  table[row sep=crcr]{%
4	80\\
8	320\\
12	480\\
16	640\\
20	800\\
24	960\\
28	1120\\
32	1280\\
};
\addlegendentry{Star (min)}

\addplot [color=mycolor7, line width=1.5pt, mark size=4.0pt, mark=triangle, mark options={solid, mycolor7}]
  table[row sep=crcr]{%
4	320.0\\
8	1280.0\\
12	1920.0\\
16	2560.0\\
20	3200.0\\
24	3840.0\\
28	4480.0\\
32	5120.0\\
};
\addlegendentry{Mesh (mean)}

\addplot [color=mycolor1, line width=1.5pt, mark size=4.0pt, mark=triangle, mark options={solid, mycolor1}, dashed]
  table[row sep=crcr]{%
4	320\\
8	1280\\
12	1920\\
16	2560\\
20	3200\\
24	3840\\
28	4480\\
32	5120\\
};
\addlegendentry{Mesh (max)}

\addplot [color=mycolor2, line width=1.5pt, mark size=4.0pt, mark=triangle, mark options={solid, mycolor2}, dotted]
  table[row sep=crcr]{%
4	320\\
8	1280\\
12	1920\\
16	2560\\
20	3200\\
24	3840\\
28	4480\\
32	5120\\
};
\addlegendentry{Mesh (min)}
\end{axis}
\end{tikzpicture}%
    \vspace{-1.cm}
  \centerline{(a) Signaling overhead versus \ac{ut} number $U$} \medskip
\end{minipage}
\hfill
\begin{minipage}[b]{0.98\linewidth}
  \centering
%
\definecolor{mycolor1}{rgb}{0.00000,0.44700,0.74100}%
\definecolor{mycolor2}{rgb}{0.85000,0.32500,0.09800}%
\definecolor{mycolor3}{rgb}{0.92900,0.69400,0.12500}%
\definecolor{mycolor4}{rgb}{0.49400,0.18400,0.55600}%
\definecolor{mycolor5}{rgb}{0.46600,0.67400,0.18800}%
\definecolor{mycolor6}{rgb}{0.30100,0.74500,0.93300}%
\definecolor{mycolor7}{rgb}{0.63500,0.07800,0.18400}%
\begin{tikzpicture}

\begin{axis}[%
width=72mm,
height=38mm,
at={(0mm, 0mm)},
scale only axis,
xtick={3, 4, 5, 6, 7, 8, 9, 10, 11, 12},
xmin=3,
xmax=12,
xlabel style={font=\color{white!15!black}, font=\footnotesize},
xlabel={LEO satellite number $S$},
ymin=0,
ymax=36000,
ylabel style={font=\color{white!15!black}, font=\footnotesize},
ylabel={Signaling overhead},
axis background/.style={fill=white},
xmajorgrids,
ymajorgrids,
yminorgrids,
]
\addplot [color=mycolor1, line width=1.5pt, mark size=4.0pt, mark=o, mark options={solid, mycolor1}]
  table[row sep=crcr]{%
3	1536.0\\
4	2048.0\\
5	2560.0\\
6	3072.0\\
7	3584.0\\
8	4096.0\\
9	4608.0\\
10	5120.0\\
11	5632.0\\
12	6144.0\\
};

\addplot [color=mycolor2, line width=1.5pt, mark size=4.0pt, mark=o, mark options={solid, mycolor2}, dashed]
  table[row sep=crcr]{%
3	1536\\
4	2048\\
5	2560\\
6	3072\\
7	3584\\
8	4096\\
9	4608\\
10	5120\\
11	5632\\
12	6144\\
};

\addplot [color=mycolor3, line width=1.5pt, mark size=4.0pt, mark=o, mark options={solid, mycolor3}, dotted]
  table[row sep=crcr]{%
3	1536\\
4	2048\\
5	2560\\
6	3072\\
7	3584\\
8	4096\\
9	4608\\
10	5120\\
11	5632\\
12	6144\\
};

\addplot [color=mycolor4, line width=1.5pt, mark size=4.0pt, mark=square, mark options={solid, mycolor4}]
  table[row sep=crcr]{%
3	1024.0\\
4	1536.0\\
5	2048.0\\
6	2560.0\\
7	3072.0\\
8	3584.0\\
9	4096.0\\
10	4608.0\\
11	5120.0\\
12	5632.0\\
};

\addplot [color=mycolor5, line width=1.5pt, mark size=4.0pt, mark=square, mark options={solid, mycolor5}, dashed]
  table[row sep=crcr]{%
3	1536\\
4	3072\\
5	5120\\
6	7680\\
7	10752\\
8	14336\\
9	18432\\
10	23040\\
11	28160\\
12	33792\\
};

\addplot [color=mycolor6, line width=1.5pt, mark size=4.0pt, mark=square, mark options={solid, mycolor6}, dotted]
  table[row sep=crcr]{%
3	768\\
4	1024\\
5	1280\\
6	1536\\
7	1792\\
8	2048\\
9	2304\\
10	2560\\
11	2816\\
12	3072\\
};

\addplot [color=mycolor7, line width=1.5pt, mark size=4.0pt, mark=triangle, mark options={solid, mycolor7}]
  table[row sep=crcr]{%
3	1536.0\\
4	3072.0\\
5	5120.0\\
6	7680.0\\
7	10752.0\\
8	14336.0\\
9	18432.0\\
10	23040.0\\
11	28160.0\\
12	33792.0\\
};

\addplot [color=mycolor1, line width=1.5pt, mark size=4.0pt, mark=triangle, mark options={solid, mycolor1}, dashed]
  table[row sep=crcr]{%
3	1536\\
4	3072\\
5	5120\\
6	7680\\
7	10752\\
8	14336\\
9	18432\\
10	23040\\
11	28160\\
12	33792\\
};

\addplot [color=mycolor2, line width=1.5pt, mark size=4.0pt, mark=triangle, mark options={solid, mycolor2}, dotted]
  table[row sep=crcr]{%
3	1536\\
4	3072\\
5	5120\\
6	7680\\
7	10752\\
8	14336\\
9	18432\\
10	23040\\
11	28160\\
12	33792\\
};
\end{axis}
\end{tikzpicture}%
    \vspace{-1.cm}
  \centerline{(b) Signaling overhead versus \ac{leo} satellite number $S$} \medskip
\end{minipage}
\vspace{-0.4cm}
\caption{Signaling overhead comparison under various schemes.}
\label{Overhead_comparison}
\vspace{-0.2cm}
\end{figure}

\subsubsection{Sum Rate}
In Figs. \ref{SR_comparison}(a)-(c), we compare the sum rate of different schemes as functions of the per-satellite power budget $P_s$, the number of antennas $N$, and the number of \ac{leo} satellites, respectively. Across all considered parameter ranges, the networked \ac{leo} cooperative beamforming schemes relying on the proposed optimization algorithms consistently and significantly outperform the closed-form baselines, namely \ac{mrt} and \ac{zf}, demonstrating the advantage of optimization-based beamforming refinement over heuristic designs.

Moreover, the decentralized schemes track the centralized upper bound nearly identically across all scenarios, while offering substantially improved scalability. The instantaneous-\ac{csi} centralized scheme curve sits above all statistical-\ac{csi} schemes throughout. Across all considered configurations, the gap to the idealized upper bound stays minor, indicating that operating with statistical \ac{csi} incurs only a modest performance loss. In contrast, the \ac{sss}-based schemes perform markedly worse than most networked schemes, underlining the importance of constellation-level cooperation. The lone exception is that \ac{sss} outperforms \ac{zf} in most cases, attributable to the wide beam footprint of satellite links inducing severe \ac{iui}: enforcing complete interference nulling via \ac{zf} overly restricts the beamforming design and substantially degrades the achievable sum rate.

\begin{figure}[t]
\centering
\begin{minipage}[b]{0.98\linewidth}
\vspace{-0.3cm}
  \centering
%
\definecolor{mycolor1}{rgb}{0.00000,0.44700,0.74100}%
\definecolor{mycolor2}{rgb}{0.85000,0.32500,0.09800}%
\definecolor{mycolor3}{rgb}{0.92900,0.69400,0.12500}%
\definecolor{mycolor4}{rgb}{0.49400,0.18400,0.55600}%
\definecolor{mycolor5}{rgb}{0.46600,0.67400,0.18800}%
\definecolor{mycolor6}{rgb}{0.30100,0.74500,0.93300}%
\definecolor{mycolor7}{rgb}{0.63500,0.07800,0.18400}%
\begin{tikzpicture}

\begin{axis}[%
width=72mm,
height=38mm,
at={(0mm, 0mm)},
scale only axis,
xtick={4, 8, 12, 16, 20, 24, 28, 32},
xmin=4,
xmax=32,
xlabel style={font=\color{white!15!black}, font=\footnotesize},
xlabel={UT number $U$},
ymode=log,
ymin=0.001,
ymax=100,
ylabel style={font=\color{white!15!black}, font=\footnotesize},
ylabel={Running time [s]},
axis background/.style={fill=white},
xmajorgrids,
ymajorgrids,
yminorgrids,
legend style={at={(-0.1,1.040)}, font=\footnotesize, anchor=south west, legend cell align=left, align=left, draw=white!15!black, legend columns = 2}
]
\addplot [color=mycolor1, line width=1.5pt, mark size=4.0pt, mark=o, mark options={solid, mycolor1}]
  table[row sep=crcr]{%
4	1.7104798957498133\\
8	9.319869098545247\\
12	15.600982898110589\\
16	21.36560210981803\\
20	27.0602297756156\\
24	34.31852020228585\\
28	40.41573381113364\\
32	47.52651438985725\\
};
\addlegendentry{Centralized}

\addplot [color=mycolor2, line width=1.5pt, mark size=4.0pt, mark=square, mark options={solid, mycolor2}]
  table[row sep=crcr]{%
4	0.0026088592665877916\\
8	0.013170387499970578\\
12	0.033264246660110075\\
16	0.06393493340001441\\
20	0.10936686659988482\\
24	0.1967550066599506\\
28	0.22784526931817792\\
32	0.299994182500086\\
};
\addlegendentry{Decentralized-Ring}

\addplot [color=mycolor3, line width=1.5pt, mark size=4.0pt, mark=triangle, mark options={solid, mycolor3}]
  table[row sep=crcr]{%
4	0.002667332692362834\\
8	0.013742199023545254\\
12	0.03664028999997148\\
16	0.07070637920041918\\
20	0.11392754601743629\\
24	0.16906708653841634\\
28	0.11147345494073643\\
32	0.22046073409093714\\
};
\addlegendentry{Decentralized-Star}

\addplot [color=mycolor4, line width=1.5pt, mark size=4.0pt, mark=diamond, mark options={solid, mycolor4}]
  table[row sep=crcr]{%
4	0.002871204175062303\\
8	0.0159341810726222\\
12	0.03142029499998898\\
16	0.06826942418003455\\
20	0.08656685606368807\\
24	0.14173958055552147\\
28	0.12530941413044827\\
32	0.17952246851119627\\
};
\addlegendentry{Decentralized-Mesh}

\addplot [color=mycolor5, line width=1.5pt, mark size=4.0pt, mark=square, mark options={solid, mycolor5}, dashed]
  table[row sep=crcr]{%
4	0.19599745093324195\\
8	0.6471446243000779\\
12	1.4680335733199898\\
16	2.8962630375004665\\
20	4.279163720799988\\
24	6.683481047500099\\
28	7.343038180161897\\
32	11.78447572249992\\
};
\addlegendentry{Decentralized-Ring (CVX)}

\addplot [color=mycolor6, line width=1.5pt, mark size=4.0pt, mark=triangle, mark options={solid, mycolor6}, dashed]
  table[row sep=crcr]{%
4	0.18040011026146552\\
8	0.6582526068588349\\
12	1.4999391716666286\\
16	2.6450725250004323\\
20	3.686305101890916\\
24	5.440892468599934\\
28	5.547888368518518\\
32	7.671978419690938\\
};
\addlegendentry{Decentralized-Star (CVX)}

\addplot [color=mycolor7, line width=1.5pt, mark size=4.0pt, mark=diamond, mark options={solid, mycolor7}, dashed]
  table[row sep=crcr]{%
4	0.31671842289997587\\
8	1.4444677613635908\\
12	1.4495480425001006\\
16	2.665937493181925\\
20	3.3265443518086535\\
24	5.470113457422283\\
28	4.772897509615443\\
32	9.069580580555582\\
};
\addlegendentry{Decentralized-Mesh (CVX)}
\end{axis}
\end{tikzpicture}%
    \vspace{-1.cm}
  \centerline{(a) Running time versus \ac{ut} number $U$} \medskip
\end{minipage}
\hfill
\begin{minipage}[b]{0.98\linewidth}
  \centering
%
\definecolor{mycolor1}{rgb}{0.00000,0.44700,0.74100}%
\definecolor{mycolor2}{rgb}{0.85000,0.32500,0.09800}%
\definecolor{mycolor3}{rgb}{0.92900,0.69400,0.12500}%
\definecolor{mycolor4}{rgb}{0.49400,0.18400,0.55600}%
\definecolor{mycolor5}{rgb}{0.46600,0.67400,0.18800}%
\definecolor{mycolor6}{rgb}{0.30100,0.74500,0.93300}%
\definecolor{mycolor7}{rgb}{0.63500,0.07800,0.18400}%
\begin{tikzpicture}

\begin{axis}[%
width=72mm,
height=38mm,
at={(0mm, 0mm)},
scale only axis,
xmin=4,
xmax=8,
xlabel style={font=\color{white!15!black}, font=\footnotesize},
xlabel={LEO satellite number $S$},
ymode=log,
ymin=0.1,
ymax=1000,
ylabel style={font=\color{white!15!black}, font=\footnotesize},
ylabel={Running time [s]},
axis background/.style={fill=white},
xmajorgrids,
ymajorgrids,
yminorgrids,
]
\addplot [color=mycolor1, line width=1.5pt, mark size=4.0pt, mark=o, mark options={solid, mycolor1}]
  table[row sep=crcr]{%
4	16.841242775999945\\
5	29.429684824428737\\
6	46.189884410714775\\
7	68.37765352971473\\
8	107.36162739062456\\
};

\addplot [color=mycolor2, line width=1.5pt, mark size=4.0pt, mark=square, mark options={solid, mycolor2}]
  table[row sep=crcr]{%
4	0.21855143866661011\\
5	0.19616875416002585\\
6	0.20766299306055633\\
7	0.18470604166672966\\
8	0.19298366746156367\\
};

\addplot [color=mycolor3, line width=1.5pt, mark size=4.0pt, mark=triangle, mark options={solid, mycolor3}]
  table[row sep=crcr]{%
4	0.1753995937497166\\
5	0.1899284560726797\\
6	0.17426932639401374\\
7	0.17383469345241748\\
8	0.1636656514196407\\
};

\addplot [color=mycolor4, line width=1.5pt, mark size=4.0pt, mark=diamond, mark options={solid, mycolor4}]
  table[row sep=crcr]{%
4	0.18658935312505492\\
5	0.1705634898221534\\
6	0.17618099846286997\\
7	0.16423768585717347\\
8	0.16096890677499687\\
};

\addplot [color=mycolor5, line width=1.5pt, mark size=4.0pt, mark=square, mark options={solid, mycolor5}, dashed]
  table[row sep=crcr]{%
4	7.426470130777791\\
5	7.907178546679933\\
6	8.747255300515187\\
7	9.270761998011862\\
8	9.692244231971086\\
};

\addplot [color=mycolor6, line width=1.5pt, mark size=4.0pt, mark=triangle, mark options={solid, mycolor6}, dashed]
  table[row sep=crcr]{%
4	7.160345322874491\\
5	7.354209813636341\\
6	7.947750050500018\\
7	8.582319012892878\\
8	9.215814626116103\\
};

\addplot [color=mycolor7, line width=1.5pt, mark size=4.0pt, mark=diamond, mark options={solid, mycolor7}, dashed]
  table[row sep=crcr]{%
4	7.188260991649986\\
5	7.369012112955614\\
6	7.947868010796271\\
7	8.73353549469845\\
8	9.574941265100005\\
};
\end{axis}
\end{tikzpicture}%
    \vspace{-1.cm}
  \centerline{(b) Running time versus \ac{leo} satellite number $S$} \medskip
\end{minipage}
\vspace{-0.4cm}
\caption{Running time comparison under various schemes.}
\label{Runtime_comparison}
\vspace{-0.2cm}
\end{figure}

\subsubsection{Signaling Overhead}

Fig. \ref{Overhead_topo} compares the per-satellite signaling overhead the proposed decentralized algorithm incurs to approach centralized performance under the three representative \ac{isl} topologies. As expected, the Mesh topology incurs the highest and uniformly distributed signaling overhead due to its fully connected structure, reflecting the cost of the most comprehensive information exchange. The Ring and Star topologies, in contrast, exhibit substantially lower overhead. Within the Star topology, however, the central satellite carries an overhead comparable to that of the Mesh topology since it serves as the information hub, while the edge satellites incur lower overhead than the Ring satellites because they communicate with only one neighbor.

Figs. \ref{Overhead_comparison}(a) and (b) further illustrate the signaling overhead as functions of the number of \acp{ut} and the number of \ac{leo} satellites, respectively. For all topologies, the overhead scales approximately linearly with $U$, consistent with the expression $|\mathcal{G}_s||\mathcal{U}_s|SU$ derived in Section~\ref{sec-overhead}, noting that $|\mathcal{U}_s|$ is bounded by the number of \acp{rfc}. With respect to $S$, the overhead scales linearly for the Ring topology and for edge satellites in the Star topology, since $|\mathcal{G}_s|$ remains constant in these cases. On average, the Star topology achieves the lowest network-wide signaling overhead, albeit at the expense of an unbalanced load concentrated at the central satellite. The Mesh topology, in contrast, exhibits polynomial growth in both maximal and average signaling overhead as $S$ increases, exposing a scalability limitation from the signaling perspective. Given that the Mesh topology achieves sum rate comparable to that of the Ring and Star topologies, these results indicate that the proposed framework can be effectively deployed over the more practical Ring or Star \ac{isl} topologies, preserving performance while significantly reducing signaling overhead and thereby demonstrating its robustness and scalability.

\subsubsection{Running Time}
In Figs. \ref{Runtime_comparison}(a) and (b), we compare the running time of three schemes: (i) the centralized scheme of Algorithm~\ref{wmmse_cent}, (ii) the original decentralized scheme of Algorithm~\ref{wmmse_cadmm_cvx}, and (iii) the proposed low-complexity decentralized scheme obtained by replacing line~7 of Algorithm~\ref{wmmse_cadmm_cvx} with Algorithm~\ref{lc_solution}, swept against $U$, $S$, and $N$. Across all considered ranges, the proposed low-complexity decentralized scheme achieves orders-of-magnitude reductions in running time compared to the centralized approach, highlighting its effectiveness for practical \ac{leo} deployment. Notably, when CVX is employed for local optimization, the decentralized runtime can even approach that of the centralized one, as predicted by Remark~\ref{decent_cvx_complex}, confirming the necessity of the low-complexity design for scalable decentralized implementation.

\section{Conclusion}\label{sec_con}
This paper studied decentralized cooperative beamforming for networked \ac{leo} satellite downlink systems enabled by \acp{isl}. We developed a topology-agnostic and fully decentralized beamforming framework that admits parallel per-satellite execution and scales efficiently to large constellations. Starting from a centralized \ac{wmmse}-based benchmark, we integrated \ac{wmmse} with \ac{cadmm} to enable decentralized optimization over arbitrary connected inter-satellite networks. By eliminating consensus-related auxiliary variables in closed form, we further derived a low-complexity yet optimal per-satellite update rule with a quasi-closed-form solution. Numerical results demonstrated that the proposed decentralized schemes closely approach centralized performance under practical inter-satellite topologies such as Ring and Star, while significantly reducing computational complexity and signaling overhead. These findings indicate that efficient constellation-level cooperation can be achieved without dense inter-satellite connectivity, making the proposed framework well suited for scalable deployment in large \ac{leo} satellite networks.

Future work will explore more realistic \ac{isl} constraints, joint scheduling-and-beamforming via mixed-integer or learning-augmented relaxations\cite{meixia2024twc}, asynchronous \ac{cadmm} extensions for dynamic graphs, robustness to residual calibration errors, and extensions to integrated sensing and communication and multi-orbit satellite systems.

\appendices

\bibliographystyle{IEEEtran}
\bibliography{IEEEabrv,mybib}

\end{document}